\documentclass[sigconf]{acmart}
\usepackage{enumitem} 
\usepackage{amsmath}
\usepackage{multirow}
\usepackage{booktabs}
\usepackage{graphicx}
\usepackage{adjustbox}
\usepackage[normalem]{ulem}
\usepackage{appendix}
\usepackage{hyperref}
\usepackage{placeins}

\newcommand{\inlinefigure}[2][1]{\adjustbox{valign=t,totalheight=#1\baselineskip,keepaspectratio}{\includegraphics{#2}}}
\setlist[itemize]{leftmargin=4mm}
\setlist[enumerate]{leftmargin=4mm}

\newcommand{\ie}{\emph{i.e., }}
\newcommand{\eg}{\emph{e.g., }}

\AtBeginDocument{%
  \providecommand\BibTeX{{%
    \normalfont B\kern-0.5em{\scshape i\kern-0.25em b}\kern-0.8em\TeX}}}

\copyrightyear{2025}
\acmYear{2025}
\setcopyright{acmlicensed}\acmConference[KDD '25]{Proceedings of the 31st ACM SIGKDD Conference on Knowledge Discovery and Data Mining V.1}{August 3--7, 2025}{Toronto, ON, Canada}
\acmBooktitle{Proceedings of the 31st ACM SIGKDD Conference on Knowledge Discovery and Data Mining V.1 (KDD '25), August 3--7, 2025, Toronto, ON, Canada}
\acmDOI{10.1145/3690624.3709255}
\acmISBN{979-8-4007-1245-6/25/08}

\begin{document}

\title{
Fine-tuning Multimodal Large Language Models for Product Bundling
}

\author{Xiaohao Liu}
\email{xiaohao.liu@u.nus.edu}
\affiliation{%
  \institution{National University of Singapore}
  \city{Singapore}
  \country{Singapore}
}

\author{Jie Wu}
\email{wujie@cuc.edu.cn}
\affiliation{%
  \institution{Communication University of China}
  \city{Beijing}
  \country{China}
}

\author{Zhulin Tao}
\email{taozhulin@gmail.com}
\affiliation{%
  \institution{Communication University of China}
  \city{Beijing}
  \country{China}
}
\authornote{Corresponding author.}

\author{Yunshan Ma}
\email{yunshan.ma@u.nus.edu}
\affiliation{%
  \institution{National University of Singapore}
  \city{Singapore}
  \country{Singapore}
}

\author{Yinwei Wei}
\email{yinwei.wei@hotmail.com}
\affiliation{%
  \institution{Shandong University}
  \city{Shangdong}
  \country{China}
}

\author{Tat-seng Chua}
\email{dcscts@nus.edu.sg}
\affiliation{%
  \institution{National University of Singapore}
  \city{Singapore}
  \country{Singapore}
}

\renewcommand{\shortauthors}{Xiaohao Liu\ et\ al.}

\begin{abstract}

Recent advances in product bundling have leveraged multimodal information through sophisticated encoders, but remain constrained by limited semantic understanding and a narrow scope of knowledge. Therefore, some attempts employ In-context Learning (ICL) to explore the potential of large language models (LLMs) for their extensive knowledge and complex reasoning abilities. However, these efforts are inadequate in understanding mulitmodal data and exploiting LLMs' knowledge for product bundling. To bridge the gap, we introduce Bundle-MLLM, a novel framework that fine-tunes LLMs through a hybrid item tokenization approach within a well-designed optimization strategy. Specifically, we integrate textual, media, and relational data into a unified tokenization, introducing a soft separation token to distinguish between textual and non-textual tokens. Additionally, a streamlined yet powerful multimodal fusion module is employed to embed all non-textual features into a single, informative token, significantly boosting efficiency. To tailor product bundling tasks for LLMs, we reformulate the task as a multiple-choice question with candidate items as options. We further propose a progressive optimization strategy that fine-tunes LLMs for disentangled objectives:   learning bundle patterns and enhancing multimodal semantic understanding specific to product bundling. Extensive experiments demonstrate that our approach outperforms a range of state-of-the-art (SOTA) methods~\footnote{Codes are available at \url{https://github.com/Xiaohao-Liu/Bundle-MLLM}}.

\end{abstract}



\maketitle

\section{Introduction}

\begin{figure}[t!]
    \centering
    \includegraphics[width=0.95\linewidth]{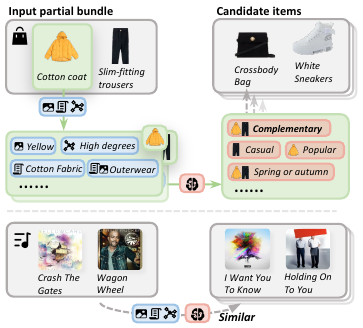}
    \vspace{-3mm}
    \caption{Illustrative example of how the multiple semantics from intrinsically multimodal data \inlinefigure{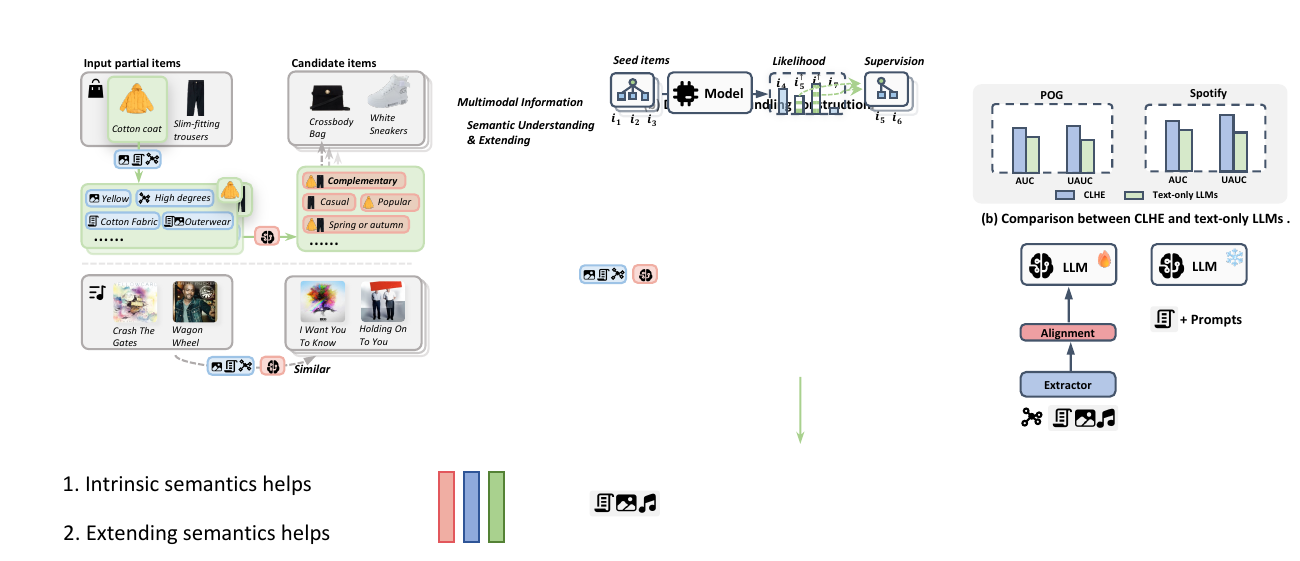} and extensive knowledge \inlinefigure{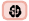} benefit the product bundling across different domains (\eg outfits or playlists).}
    \label{fig:top_fig}
    \vspace{-4mm}
\end{figure}

Product bundling is a prevailing marketing strategy where multiple products are combined into an unified item set (\ie bundle) and offered together at a discounted price~\cite{listRec, packageRec}. 
Such strategy can significantly boost sales by encouraging customers to purchase more items together, thereby increasing the average transaction value. 
It also enhances customer satisfaction by offering perceived savings and convenience, potentially leading to greater customer loyalty and repeat business.
In practice, product bundling necessitates multi-faceted information and domain knowledge, which is labor intensive and resources costly~\cite{bakos2000bundling}.
To this end, automatic product bundling has attracted considerable interests, heralding a novel avenue for creating appealing bundles~\cite{BGN, BGGN, CLHE}.

The development of automatic production bundling has clearly evolved, shifting from mere co-occurrence-based methods to more sophisticated approaches that integrate multimodal information.
Traditional methods predict the remaining items in a bundle based on the co-occurrence relationships of items within existing bundles~\cite{BundleNet, BGGN}. 
Modern methods incorporate multimodal information to enhance item representations, offering a stronger rationale for bundling at the content level.~\cite{POG, CLHE}. 
However, these newer methods are constrained by their limited semantic understanding and the narrow scope of knowledge within a compact model.
Semantic understanding helps construct the semantic context for items across multiple modalities. For example, a cotton coat can be characterized by various semantic attributes derived from different multimodal sources, such as \emph{yellow} for its color and \emph{cotton fabric} for its material.
Furthermore, extensive knowledge takes into account the nuanced needs of the input partial bundle, encapsulating complex bundling strategies.
For instance, as shown in Figure~\ref{fig:top_fig}, a cotton coat paired with slim-fitting trousers might suggest a \emph{complementary} need, associated with \emph{spring or autumn} for seasonality, while two rock songs might indicate a \emph{similar} need.
Despite their importance, these aspects were largely overlooked until the advent of large language models (LLMs), which are capable of understanding fine-grained semantics from tokens and are trained on diverse corpora.

Recognizing the potential of adapting LLMs for product bundling, AICL~\cite{bundleICL} employs in-context learning (ICL) to retrieve relevant bundle samples for effective prompting. 
However, such approach falls short in two key areas: they fail to leverage multimodal features (\ie media content and relational data) and cannot fully exploit LLMs specifically for bundling tasks.
Multimodal features are crucial for enhancing item representation in product bundling. For instance, media content (\ie acoustic or visual media) can compensate for the limitations of textual descriptions by providing richer, more nuanced item representations. 
Similarly, relational data (\ie user-item interactions and bundle-item affiliations) introduces global and heterogeneous correlations among items, extending the limited context that can be captured within prompts.
Moreover, fine-tuning LLMs is essential to comprehend multimodal features and invoke bundle-specific knowledge, therefore achieving accurate and relevant bundling outputs.
However, recent multimodal LLMs (MLLMs), such as vision-text~\cite{minigpt4, llava} or audio-text models~\cite{qwen_audio, speechGPT}, rely heavily on extensive data to align different modalities, primarily for captioning tasks. This approach incurs significant resource costs and falls short in addressing the diverse modalities (\eg relational data) and the multiple items needed for effective product bundling.
Thus, the challenge of fine-tuning \emph{bundle-specific} LLMs to incorporate \emph{multimodal information} remains unresolved and demands further research.

To this end, we propose a novel multimodal large language framework for product bundling, termed Bundle-MLLM.
This framework introduces hybrid item tokenization, enabling the model to process not only descriptive text but also diverse item modalities. 
Specifically, media features are extracted using foundation encoders like BLIP2~\cite{blip2} for visual media and CLAP~\cite{clap} for acoustic media, while relational features are derived from a pre-trained collaborative filtering method (\eg LightGCN~\cite{Lightgcn}).
Notably, all non-textual features are embedded into a single token per item through a streamlined yet effective fusion module, designed to maximize the interplay among different modalities and boost efficiency. 
To tailor a bundle-specific MLLM, we fine-tune it with a few samples in a progressive optimization strategy to invoke its extensive knowledge to answer product bundling.  
This strategy begins with updating the LoRA weights using text-only prompts, followed by training the multimodal fusion module and a projector to fit the task-aligned LLMs. 
By employing this strategy, we can maximize the synergy between bundle pattern learning and multimodal semantic understanding in Bundle-MLLM.
Our design is both simple and efficient, demonstrating significant bundling efficacy that facilitates practical deployment and further advances domain development.
Extensive experiments on four datasets across two domains demonstrate that our method outperforms multiple leading methods, including the traditional state-of-the-art (SOTA) method (\ie CLHE~\cite{CLHE}) and GPT-4~\cite{gpt4}). 
Furthermore, ablation studies and model analyses confirm the effectiveness of key modules and underscore the critical properties of the proposed model, such as its anti-cold-start capabilities.
We summarize the key contributions of this work as follows:
\begin{itemize}
    \item~We pioneer the adaptation of MLLMs for product bundling, addressing prevailing limitations such as inferior semantic understanding and restricted scope of knowledge.
    \item~We introduce Bundle-MLLM, which incorporates multimodal features through hybrid item tokenization and utilizes progressive optimization to achieve optimal bundling performance.
    \item~Our method outperforms various leading methods on four datasets across two application domains, and diverse studies demonstrate the various merits of our method.
\end{itemize}

\section{Problem formulation}

Given a set of items $\mathcal{I}=\{i_1, i_2, \dots, i_N\}$ and a set of bundles $\mathcal{B} = \{b_1, b_2, \dots, b_M\}$, where $N$ ($M$) denotes the number of items (bundles), \ie $N = |\mathcal{I}|$, $M = |\mathcal{B}|$, 
we define a bundle as a set of items denoted as $b = \{i_1, i_2, \dots, i_n\}$, where $n = |b|$ is the size of the bundle.
Given the text token set $\mathcal{T}$ and media set $\mathcal{M}$, each item has a textual input $t_i \in \mathcal{T}^{l_i}$ (where $l_i$ is the token length of $t_i$), which can be its title, description, or metadata, and a media input $m_i \in \mathcal{M}$, which can be an image or audio of the item. 
In addition, for the items that have been online for a while, we have collected some item-level user feedback data, which is denoted as a user-item interaction matrix $\mathbf{X}_{O\times N}=\{x_{u,i} \mid u \in \mathcal{U}, i \in \mathcal{I}\}$, where $\mathcal{U} = \{u_1, u_2, \dots, u_O\}$ is the user set with a size of $O = |\mathcal{U}|$.
Given the defined bundle-item affiliation matrix $\mathbf{Y}_{M\times N}=\{x_{b,i} \mid b \in \mathcal{B}, i \in \mathcal{I}\}$,
the product bundling aims to predict unseen bundles, denoted as $\mathcal{\bar{B}}=\{b_{M+1}, b_{M+2}, \dots, b_{M+\bar{M}}\}$, where $\bar{M} = |\mathcal{\bar{B}}|$. We aim to obtain a model for an unseen bundle $b \in \mathcal{\bar{B}}$ such that, when given a few seed items $\bar{b}_s \subset b$, \ie, the partial bundle, the model can predict the missing items $b \setminus \bar{b}_s$, thereby constructing the entire bundle.

\section{Methodology}

\begin{figure*}
    \centering
    \includegraphics[width=0.99\linewidth]{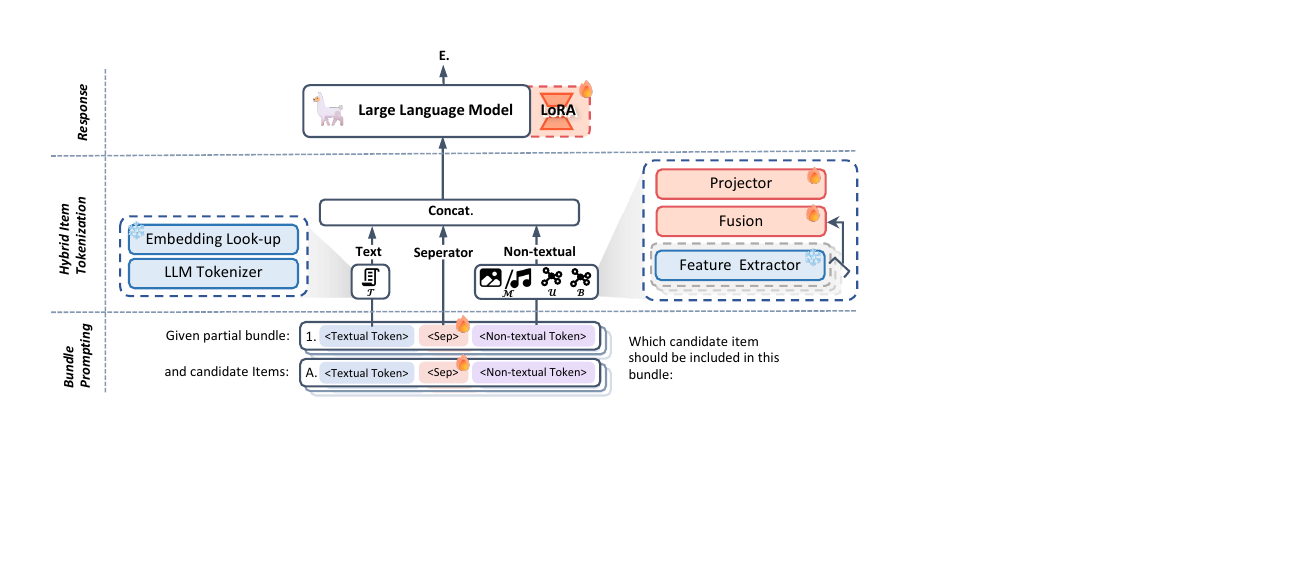}
    \caption{The overall framework of Bundle-MLLM, which incorporates multiple data formats, including textual, visual/acoustic, and relational features (\ie user-item interactions and bundle-item affiliations).
    These heterogeneous features are extracted from various foundation encoders and embedded into a single multimodal token via a trainable fusion module, followed by a projector to align with the LLM space.}
    \label{fig:framework}
\end{figure*}

To achieve bundle capability for MLLMs and uncover the semantics from multiple data formats (i.e., textual, visual/acoustic, and relational data), we propose Bundle-MLLM, as illustrated in Figure~\ref{fig:framework}. 
Bundle-MLLM processes textual descriptions and heterogeneous non-textual features through hybrid item tokenization, aggregating extracted multimodal features into a single yet informative token embedding. 
This approach also alleviates the computational burden associated with the autoregressive inference of LLMs.
The bundle prompting helps organize the hierarchy and align tasks in a manner that LLMs prefer. Additionally, we employ progressive optimization to adapt LLMs from bundle pattern learning to multimodal semantic understanding, maximizing the synergy of modules.

\subsection{Hybrid Item Tokenization}

To incorporate multiple data formats, including textual descriptions, media content, user-item interactions, and bundle-item affiliations, we introduce hybrid item tokenization.

\noindent\textbf{Textual Tokenization.}
Item textual features serve as a natural format for LLM processing, involving titles or categorical tags that are informative for direct semantic extraction.
We leverage the inherent tokenizer and stored embedding table to transform textual input into latent embeddings.
This process can be formally represented as:
\begin{equation}
    z^{\text{t}}_{i;p}  = \mathcal{E}^{\text{t}}(t_{i;p}), \mathcal{E}^{\text{t}} :=  \text{Tokenizer} \circ \text{LLM-Emb},
    \label{eq:llm_token}
\end{equation}
where $p\in[l_i]$ is the position of the token within input $t_{i}$. 
Here $\mathcal{E}^{\text{t}}$ is denoted as the composition of $\text{Tokenizer}$ and $\text{LLM-Emb}$. 
The textual input of item $i$ can be represented as a concatenation of an ordered series of latent embeddings, formally, 
$z^{\text{t}} = [z^{\text{t}}_{i;1}, z^{\text{t}}_{i;2}, \dots, z^{\text{t}}_{i;l_i}]$.
Note that all textual inputs, except the item description, such as task instructions, utilize the same procedure to obtain embedding that is fed into the LLM.

\noindent\textbf{Heterogeneous Feature Extraction.}
The wealth of heterogeneous multimodal information enhances the intrinsic semantics, enriching the sole textual effect.  
To integrate heterogeneous modalities (\ie media content, user-item interactions and pre-defined bundle-item affiliations), we employ corresponding encoders to effectively extract informative representations. 
Note that we do not include textual information in this extraction framework since LLMs are inherently well-performing textual extractors.
With the advancement of foundation models, we utilize the pre-trained model BLIP2~\cite{blip2} for visual content and CLAP~\cite{clap} for acoustic content to yield the representations $z^{\text{m}}_{i} \in \mathbb{R}^{d_m}$ for item $i$.
Formally, 
\begin{equation}
    z^{\text{m}}_{i} = \mathcal{E}^{\text{m}}(m_i; \Theta_{\text{m}}),\quad \mathcal{E}^{\text{m}} \in \{\text{BLIP2}, \text{CLAP}\},
    \label{eq:pre_traineder}
\end{equation}
where $\mathcal{E}^{\text{m}}$, equipped with its parameter $\Theta_{\text{m}}$, depends on the specific dataset.

For the relational data (\ie $\mathbf{X}$ and $\mathbf{Y}$), we adopt the well-recognized CF method, LightGCN~\cite{Lightgcn}, to obtain the item representation $z^{\text{x}}_{i} \in \mathbb{R}^{d_x}$/$z^{\text{y}}_{i} \in \mathbb{R}^{d_y}$ from various-order interactions/affiliations.  
Specifically, we initialize the user embeddings $\{e_u^\text{x} \in \mathbb{R}^{\tilde{d}}; u \in \mathcal{U}\}$ and item embeddings $\{e_i^\text{x} \in \mathbb{R}^{\tilde{d}}; i \in \mathcal{I}\}$.
We then train the LightGCN model, which utilizes a message-passing mechanism, over the bipartite graph, defined as follows:
\begin{equation}
    e_u^\text{x;k} = \sum_{i\in \mathcal{N}_u} \frac{e_u^\text{x;k-1}}{\sqrt{|\mathcal{N}_u|}\sqrt{|\mathcal{N}_i|}},\quad
    e_i^\text{x;k} = \sum_{u\in \mathcal{N}_i} \frac{e_i^\text{x;k-1}}{\sqrt{|\mathcal{N}_i|}\sqrt{|\mathcal{N}_u|}},
    \label{eq:message-passing}
\end{equation}
where $k$ denotes the layer indicator in the LightGCN process. By default, we set $e_u^\text{x;0} = e_u^\text{x}$ and $e_i^\text{x;0} = e_i^\text{x}$.
$\mathcal{N}_u$ and $\mathcal{N}_i$ denote the neighbors of the user $u$ and item $i$, respectively. 
Finally, we store the user-level item representation by aggregating the item representations over $K$ layers of propagation, represented as:
\begin{equation}
    z^{\text{x}}_{i} = \frac{1}{K}\sum_{k\in[K]} e_i^\text{x;k}.
\end{equation}
Similarly, for the bundle-item affiliations, we employ the same training procedure while using the independently initialized parameters $\{e_b^\text{y} \in \mathbb{R}^{\tilde{d}}; b \in \mathcal{B}\}$ and $\{e_i^\text{y} \in \mathbb{R}^{\tilde{d}}; i \in \mathcal{I}\}$.
Through the message-passing mechanism, similar to Equation~\ref{eq:message-passing}, we obtain the bundle-level item representation: $z^{\text{y}}_{i} = \frac{1}{K}\sum_{k\in[K]} e_i^\text{y;k}$.

\noindent\textbf{Non-textual Tokenization.}
There is a naive implementation to incorporate the aforementioned non-textual features into organized tokens:

\begin{equation}
    z_{i} = [ 
    \underbrace{\mathcal{E}^\text{t}(t_\text{m}), z^\text{m}_{i\phantom{y}}}_{\text{media}}, 
    \underbrace{\mathcal{E}^\text{t}(t_\text{x}), z^\text{x}_{i\phantom{y}}}_{\text{user-level}}, 
    \underbrace{\mathcal{E}^\text{t}(t_\text{y}), z^\text{y}_{i\phantom{y}}}_{\text{bundle-level}}
    ].
    \label{eq:textual_prompt}
\end{equation}
where $t_\text{m}$, $t_\text{x}$, and $t_\text{y}$ represent the textual indicators for each unimodal feature (\eg $t_m := \text{"media token: "}$) to organize them and guide the LLM in comprehending the subsequent latent token. However, this approach requires a sophisticated design and might be misled by the textual indicators.
However, such approach requires a sophisticated design, even being misled by the textual indicators. 
We propose to explicitly manifest the interplay of different modalities as follows:
\begin{equation}
    z^{\text{R}}_{i} = f(z^{\text{r}}_{i}| \text{r}\in \{\text{m}, \text{x}, \text{y}\}),
    \label{eq:fusion_modality}
\end{equation}
where $f(\cdot)$ is implemented with self-attention, which has demonstrated efficacy in previous work for the product bundling~\cite{CLHE}. 
Formally,
\begin{equation}
\begin{aligned}
    \mathbf{A}^{k'}_{i} &= \frac{1}{\sqrt{d}}\mathbf{R}^{k'-1}_i\mathbf{W_Q}^{k'}(\mathbf{R}^{k'-1}_i\mathbf{W_K}^{k'})^\top, \\
    & \mathbf{R}_i^{0} = [z^{\text{m}}_i \mathbf{W}_{m},
    z^{\text{x}}_i \mathbf{W}_{x},
    z^{\text{y}}_i \mathbf{W}_{y}].
\end{aligned}
\end{equation}
For dimensional consistency, we use the transformation matrix $\mathbf{W}_{m} \in \mathbb{R}^{d_m \times d}$ and $\mathbf{W}{x}, \mathbf{W}_{y} \in \mathbb{R}^{\tilde{d} \times d}$ for the media content feature and the other two relational features. 
We then concatenate all three features as the initial input to the first-layer self-attention, denoted as $\mathbf{R}i^{0} \in \mathbb{R}^{3 \times \tilde{d}}$. 
Here, $\mathbf{W_Q}^{k'}, \mathbf{W_K}^{k'} \in \mathbb{R}^{d \times d}$ are the trainable parameters to project the input feature embeddings into the query and key spaces;
$\mathbf{A}^{k'}{i}$ is the attention matrix among the three modalities inprinR layer $k'$, followed by a softmax function for normalized attention scores to obtain new representations at the next layer, shown as:
\begin{equation}
    \mathbf{R}^{k'}_i = \text{softmax}(\mathbf{A}^{k'}_{i})\mathbf{R}^{k'-1}_i.
\end{equation}
Iteratively, the final representation $\mathbf{R}^{K'}_i$ undergoes average pooling, paired with a trainable projector, denoted as:
\begin{equation}
    z^{\text{R}}_i = \mathcal{E}^{R}(\mathbf{R}^{K'}_i),\quad
    \mathcal{E}^{R}:=  \text{average} \circ p,
    \label{eq:fuse_modal}
\end{equation}
where the projector $p(\cdot)$ is implemented by a multilayer perceptron (MLP); $\mathcal{E}^{R}$ is the composition of average pooling and projection applied to $\mathbf{R}^{K'}_i$.

\noindent\textbf{Hybrid Tokenization}
To construct complete item tokens, we utilize a straightforward method that concatenates the aforementioned latent embeddings in a structured manner, denoted as:

\begin{equation}
    z_{i} = [ 
    \underbrace{z^{\text{t}}_{i;1}, z^{\text{t}}_{i;2}, \dots, z^{\text{t}}_{i;l_i}}_{\text{<Textual Token>}} ,
    \underbrace{z^{\text{sep}}_{\phantom{i;l_i}}}_{\text{<Sep>}},
    \underbrace{z^{\text{R}}_{i\phantom{;l_i}}}_{\text{<Non-textual Token>}}
    ].
    \label{eq:fusion_token}
\end{equation}
Notably, we introduce a separate token <Sep> to distinguish the textual token and the non-textual token. 
We are inspired by the soft-prompting~\cite{prompt-tuning, prompt-tuning-newsrec} technique to introduce a learnable token. Compared to the textual hard separator, this soft token is more flexible and especially effective for datasets where the discrepancy between textual and multimodal features is significant (\eg Spotify). We also consider the direct cancellation of the separator as an alternative for datasets where the textual description is well-categorized into different semantics, see the empirical evidence in Section  \ref{sec:fusion_seperator}.

\subsection{Bundle Prompting}
To effectively integrate diverse information from items within a product bundling task for LLM inquiry, we reconstruct the traditional bundle retrieval method into a multiple-choice format. This transformation is illustrated as follows:
\begin{figure}[h]
    \centering
    \vspace{-3mm}
\includegraphics[width=0.99\linewidth]{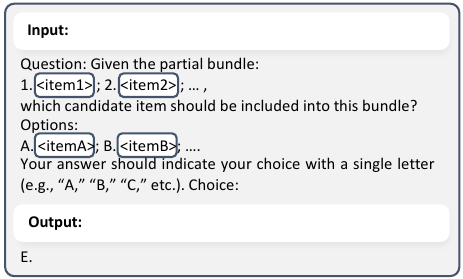}
    \label{fig:prompt}
    \vspace{-3mm}
\end{figure}

In the input seed bundle, numerical indicators are utilized to identify items, while alphabetical indicators are assigned to candidate items to serve as options within the multiple-choice framework.
Adhering to a widely accepted protocol, we instruct the LLMs to provide responses in the form of the corresponding option letter. 
In the embedding phase, these item placeholders are replaced by the aforementioned hybrid tokens (\ie $z_{i}$).

\subsection{Progressive Optimization}
\label{sec:optimization}
To achieve effective product bundling capability, we employ a progressive optimization strategy. 
For vanilla LLMs, a one-step optimization fails to balance bundling ability and multimodal semantic understanding, and even undermine each other from initialized suboptimal states. To address this, we disentangle the complete optimization objective into two phases: first, learning the bundle patterns of LLMs to fit the task, and then adapting the semantic understanding of multimodal features to fit the task-optimized LLMs.

\noindent\textbf{Bundle Pattern Learning.} 
We adopt Parameter Efficient Fine-Tuning (PEFT) for efficiently optimizing the LLMs with a smaller set of parameters. This approach significantly reduces computational requirements while still achieving commendable performance. Specifically, we choose LoRA~\cite{LoRA}, a typical PEFT algorithm, which keeps the LLM weights frozen and decomposes the updating weights into trainable low-rank matrices.
Relying on the above bundle prompting, we can optimize the parameters of LoRA as follows:
\begin{equation}
    \max_{\Theta_{\text{LoRA}}} \sum_{s\in\mathcal{S}} \sum_{\tau\in[|t^{\text{s}}|]} 
    \log 
    p(t^{\text{s}}_{\tau} | \bar{z}^{\text{s}}, t^{\text{s}}_{< \tau};\Phi,\Delta\Phi(\Theta_{\text{LoRA}})),
    \label{eq:s1}
\end{equation}
where $\Phi$ and $\Delta\Phi(\Theta_{\text{LoRA}})$ represent the frozen LLM parameters and trainable LoRA parameters, respectively. 
For each sample $s \in \mathcal{S}$, we construct the latent embeddings followed by the bundle prompting, omitting multimodal tokens, denoted as $\bar{z}^{\text{s}}$. 
And the output tokens corresponding to the correct option, is denoted as $t^{\text{s}}$. 
The variable $\tau$ represents the position of the current output token.
To avoid parameter corruption, we remove the initialized separate token and multimodal tokens (\ie $z^{\text{sep}}$ and $z^{\text{R}}_i$). 
This stage is inspired by recent instruction learning approaches, aiming to activate the bundle pattern hidden in the pre-built LLMs without diversifying its textual comprehension.

\noindent\textbf{Multimodal Semantic Understanding.}
After obtaining the well-tasked LLMs, we aim to align the fused multimodal token to the semantic space, which can be accepted for reasoning. 
To this end, we freeze the parameters of the LLM and LoRA converged from the above stage while incorporating the multimodal token processing module for optimization as follows:

\begin{equation}
    \max_{\Theta_{\text{F}}} 
    \sum_{s\in\mathcal{S}} \sum_{\tau\in[|t^{\text{s}}|]} 
    \log 
    p(t^{\text{s}}_{\tau} | z^{\text{s}}, t^{\text{s}}_{< \tau};\Phi,\Delta \Phi(\tilde{\Theta}_{\text{LoRA}}), \Theta_{\text{F}}), 
    \label{eq:s2}
\end{equation}
where $\Delta \Phi(\tilde{\Theta}_{\text{LoRA}})$ represents the optimal parameters of LoRA, and $\Theta_{\text{F}}$ denotes the parameters of the multimodal token processing module, which involves trainable multimodal fusion and projection. Note that we use $z^{\text{s}}$ to denote the complete bundle prompt for sample $s$.

\section{Experiments}

\begin{table*}
\centering
\caption{The statistics of the four datasets on two different domains.}
\vspace{-2mm}
\label{tab:dataset}
\begin{tabular}{lccccccccc} 
\toprule
\textbf{Dataset}         & \textbf{\#U} & \textbf{\#I} & \textbf{\#B} & \textbf{\#B-I} & \textbf{\#U-I} & \textbf{\#Avg.I/B} & \textbf{\#Avg.B/I} & \textbf{\#Avg.I/U} & \textbf{\#Avg.U/I}  \\ 
\midrule
\textbf{POG}             & 17,449       & 48,676       & 20,000       & 72,224         & 237,519        & 3.61               & 1.48               & 13.61              & 4.88                \\
\textbf{POG\_dense}      & 2,311,431    & 31,217       & 29,686       & 105,775        & 6,345,137      & 3.56               & 3.39               & 2.75               & 203.26              \\
\textbf{Spotify}         & 118,994      & 254,155      & 20,000       & 1,268,716      & 36,244,806     & 63.44              & 4.99               & 304.59             & 142.61              \\
\textbf{Spotify\_sparse} & 118,899      & 213,325      & 12,486       & 549,900        & 32,890,315     & 44.04              & 2.58               & 276.62             & 154.18              \\
\bottomrule
\end{tabular}
\end{table*}

\begin{table*}
\centering
\caption{The results of Bundle-MLLM compared with conventional product bundling methods and LLM-based approaches. "Rel.Imp." denotes the relative improvement of Bundle-MLLM compared to the strongest baselines. Bold and underlined indicate the best and the second-best performance.}
\label{tab:main}
\begin{tabular}{l|cc|cc|cc|cc} 
\toprule
\multirow{2}{*}{\textbf{Model}} & \multicolumn{2}{c|}{\textbf{POG}} & \multicolumn{2}{c|}{\textbf{POG\_dense}} & \multicolumn{2}{c|}{\textbf{Spotify}} & \multicolumn{2}{c}{\textbf{Spotify\_sparse}}  \\
                                & HitRate@1       & ValidRatio      & HitRate@1       & ValidRatio             & HitRate@1       & ValidRatio          & HitRate@1       & ValidRatio                  \\ 
\midrule
\textbf{MultiVAE~\cite{multiVAE}}               & 0.2786          & 1.0000          & 0.4492          & 1.0000                 & 0.5469          & 1.0000              & 0.5313          & 1.0000                      \\
\textbf{BiLSTM~\cite{BiLSTM}}                 & 0.2924          & 1.0000          & 0.4297          & 1.0000                 & 0.7201          & 1.0000              & 0.6250          & 1.0000                      \\
\textbf{UHBR~\cite{HyperGraph}}             & 0.2462          & 1.0000          & 0.5898          & 1.0000                 & 0.7434          & 1.0000              & 0.6953          & 1.0000                      \\
\textbf{CLHE~\cite{CLHE}}                   & \uline{0.3477}  & 1.0000          & \uline{0.6367}  & 1.0000                 & \uline{0.7569}  & 1.0000              & \uline{0.7227}  & 1.0000                      \\ 
\midrule
\textbf{Llama2~\cite{llama2}}                 & 0.0960         & 0.4883          & 0.0719          & 0.3906                 & 0.1998          & 0.9336              & 0.2395          & 0.9053                      \\
\textbf{ +ICL}                  & 0.0988         & 0.9492          & 0.0920          & 0.9766                 & 0.2735          & 0.9570              & 0.3108          & 0.9805                      \\
\textbf{GPT4~\cite{gpt4}}                   & 0.1786         & 0.9844          & 0.1929          & 0.9922                 & 0.6389          & 0.9843              & 0.6667          & 0.9844                      \\
\textbf{ +ICL}                  & 0.1850         & 0.9922          & 0.2196          & 0.9961                 & 0.6640          & 0.9883              & 0.7008          & 0.9922                      \\ 
\midrule
\textbf{Bundle-MLLM (ours)}                   & \textbf{0.4131} & 1.0000          & \textbf{0.6675} & 1.0000                 & \textbf{0.8242} & 1.0000              & \textbf{0.7959} & 1.0000                      \\
\textbf{Rel.Imp.}               & 18.81\%         & -               & 4.84\%          & -                      & 8.89\%          & -                   & 10.13\%         & -                           \\
\bottomrule
\end{tabular}
\end{table*}

To evaluate our proposed framework from different perspectives, we conduct experiments on four well-established datasets across two domains. 
SOTA conventional methods and advanced LLMs serve as our baselines for a comprehensive comparison. 
Furthermore, ablation studies and model studies further substantiate our claim that the proposed framework is both reasonable and effective.
To validate the superiority of our framework, we will address the following research questions:

\begin{itemize}
\item~\textbf{RQ1}: How does Bundle-MLLM perform compared to conventional product bundling models and LLM-based methods?
\item~\textbf{RQ2}: How does Bundle-MLLM perform when utilizing different modalities? Are our token processing and optimization strategies superior and beneficial?
\item~\textbf{RQ3}: How does Bundle-MLLM perform across different settings, including varying candidate sizes, cold-setting scenarios, and during the training and inference phases?
\end{itemize}

\subsection{Experimental Settings}

\subsubsection{Datasets}
Following previous work on product bundling, we evaluate our model on four established datasets across different domains, each with distinct characteristics~\cite{CLHE}, as shown in Table~\ref{tab:dataset}. 
Specifically, POG~\cite{POG} is commonly used for fashion outfits, while Spotify~\footnote{https://open.spotify.com/} is derived from music playlists. POG\_dense is a variant of POG with denser user feedback, whereas Spotify\_sparse features a smaller average bundle size than Spotify. 
Adhering to the typical setting~\cite{CLHE}, we randomly split all the bundles in each dataset into training, validation, and testing sets in an 8:1:1 ratio. The validation datasets are employed to evaluate performance and find the optimal hyperparameters.

\subsubsection{Evaluation Protocols}
\label{sec:evaluation_protocols}
For each bundle, we randomly select 10 items to construct the candidate set, which includes one positive item.
The comparison with different numbers of candidate items is detailed in Section ~\ref{sec:candidates}. 
We use HitRate@1 as the main metric to evaluate whether the models can correctly predict the positive item from the candidates. 
A higher HitRate@1 indicates a greater capability for accurate product bundling.
To ensure a fair comparison, we implement all baselines with the same number of candidates and the same evaluation algorithm. Additionally, we utilize the ValidRatio to evaluate the validity of responses from generative predictions~\cite{Llara}. 
This metric addresses potential invalid responses, such as nonsensical words or options, by quantifying the proportion of valid responses (\ie correct options within the candidate set) across all sequences.

\subsubsection{Baselines}
To comprehensively evaluate our proposed method, we compare it with two groups of methods: conventional product bundling methods and LLM-based methods. CLHE serves as the SOTA baseline, while GPT-4 paired with ICL stands as the strongest LLM-based baseline. The following is a specific description of these baselines:

\begin{itemize}
    \item \textbf{MultiVAE}~\cite{multiVAE} follows the paradigm of variational autoencoders, transforming discrete item interactions into latent representations.
    \item \textbf{BiLSTM}~\cite{BiLSTM} leverages the bi-directional LSTM technique to process bundle sequences and learn the bundle representations.
    \item \textbf{UHBR}~\cite{HyperGraph} constructs a hypergraph within bundles and employs a GCN model to obtain the bundle representations.
    \item \textbf{CLHE}~\cite{CLHE} proposes using a hierarchical encoder paired with contrastive learning to derive final representations.
    \item \textbf{Llama2}~\cite{llama2} is a well-known open-source LLM released by Meta, utilized for instruction following with pre-built parameters.
    \item \textbf{GPT-4}~\cite{gpt4} is an advanced LLM developed by OpenAI, excelling in various tasks, and is invoked via API.
    \item \textbf{ICL}~\cite{bundleICL} directly queries original LLMs through designed language prompts with several samples. We implement this technique with both Llama2 and GPT-4.
\end{itemize}
See more implementation details in Appendix~\ref{appendix:implementation details}.

\subsection{Performance Comparison (RQ1)}
To address RQ1, we conduct a comparison with extensive baselines following the aforementioned evaluation protocol to demonstrate the superiority of Bundle-MLLM, as illustrated as Table~\ref{tab:main}.
Our observations are as follows:
\begin{itemize}
    \item As expected, Bundle-MLLM outperforms all baselines by a significant margin across four datasets. 
    Specifically, it achieves up to an 18.81\% relative performance improvement \emph{w.r.t.} HitRate@1, demonstrating its effective bundle capability with multimodal semantic understanding of LLMs.
    \item Analyzing the results across four datasets, we find that the performance improvement on POG\_dense is marginal compared to the others. 
    We attribute this to the sufficient data provided, such as significant dense user-item interactions and bundle-item affiliations, which facilitate bundle/item representation learning for conventional methods (\ie CLHE), thus narrowing the performance gap with our proposed method.
    \item CLHE serves as the second strongest method by designing hierarchical learning on multimodal features and bundle representations, paired with contrastive learning. 
    Its emphasis on integrating multiple data formats underlines the rationale of our work, which explores multimodal semantics with extensive knowledge.
    \item GPT-4 achieves the best performance among the LLM-based methods, showcasing its complex reasoning capabilities. 
    Equipped with ICL, both GPT-4 and Llama2 gain certain improvements but still perform lower than conventional methods. 
    \item Bundle-MLLM also achieves a high validity ratio of 100\% across all datasets, illustrating the model's instruction-following abilities when predicting the remaining items.
    Note that all the LLM-based methods might generate invalid responses. Notably, Llama2 serves as our LLM backbone and demonstrates significant improvements of Bundle-MLLM in the validity ratio.
    Remarkably, Bundle-MLLM's significant improvement in valid ratios can be attributed to its instruction-tuning on activating bundle patterns for product bundling.
\end{itemize}

\subsection{Ablation Study (RQ2)}
To clarify RQ2, we conduct experiments to evaluate the impact of different modalities and their interplays on Bundle-MLLM. 
Moreover, to enhance the significance of our proposed hybrid item tokenization, various token processing strategies and separators are tested. 
Additionally, we delve into the development of learning strategies to justify the superiority of our proposed progressive optimization.

\subsubsection{Different modalities}

Here, we showcase three types of modality combinations to analyze their effects:
\begin{enumerate}
\item \textbf{Unimodality}: In this case, we only keep one single modality data, such as \textbf{Text}, \textbf{Media}, \textbf{UI}, and \textbf{BI}. \textbf{Text} retains the textual descriptions of items, while the others are transformed with projectors to align with the LLM space.

\item \textbf{Bimodality}: Textual description serves as the main component, paired with the other three types of modality features, denoted as \textbf{Text + Media}, \textbf{Text + UI}, and \textbf{Text + BI}.

\item \textbf{Multimodality}: We directly use \textbf{Bundle-MLLM} for comparison since its hybrid item tokenization involves all four modalities.
\end{enumerate}

The comparative results are illustrated in Table~\ref{tab:modalities}. We have the following observations.
Specifically, \textbf{Bundle-MLLM} achieves the best performance by utilizing multiple data formats, demonstrating the informative features explored by our proposed method for . 
Within the unimodality group, there is a distinct relationship of \textbf{Text} > \textbf{BI} > \textbf{Media} > \textbf{UI} for POG, Spotify, and Spotify\_sparse, while \textbf{BI} outperforms others in POG\_dense. We attribute this to the unique statistics of dense bundle-item affiliations, similar to \textbf{UI}, which also shows moderated results with sufficient user-item interactions.
Obviously, different bundle datasets have diverse emphases on different modalities, informing an attentive method to effectively integrate these data. 
The results of the interplay between textual description and other modalities (\ie bimodality) highlight the same conclusion that the direct combination of different modalities leads to inferior results. 
These observations not only clarify the motivation of our proposed nuanced multimodal process but also demonstrate the superiority and effectiveness of Bundle-MLLM.

\begin{table}
\centering
\renewcommand{\arraystretch}{1.1}
\setlength{\tabcolsep}{0.9mm}
\caption{The performance comparison \textit{w.r.t.} different modalities and their interplays.}
\vspace{-2mm}
\label{tab:modalities}
\begin{tabular}{l|c|c|c|c} 
\toprule
\textbf{Setting}            & \textbf{POG}    & \textbf{POG\_dense} & \textbf{Spotify} & \textbf{Spotify\_sparse}  \\ 
\midrule
\textbf{Text}               & 0.3759          & 0.4492              & 0.7477           & 0.7620                    \\
\textbf{Media}              & 0.0973          & 0.3932              & 0.4714           & 0.4689                    \\
\textbf{UI}                 & 0.0967          & 0.3890              & 0.1001           & 0.1027                    \\
\textbf{BI}                 & 0.2526          & 0.6128              & 0.5993           & 0.5703                    \\ 
\midrule
\textbf{Text + Media}       & 0.3631          & 0.6110              & 0.7528           & 0.7489                    \\
\textbf{Text + UI}          & 0.3895          & 0.5995              & 0.7492           & 0.7731                    \\
\textbf{\textbf{Text + BI}} & 0.2987          & 0.6243              & 0.8029           & 0.7772                    \\ 
\midrule
\textbf{Bundle-MLLM}         & \textbf{0.4131} & \textbf{0.6675}     & \textbf{0.8242}  & \textbf{0.7959}           \\
\bottomrule
\end{tabular}
\end{table}

\subsubsection{The impact of multimodal token process \& separator}
\label{sec:fusion_seperator}

To shed light on how to process hybrid tokens, we now delve into the influence of diverse multimodal token processes and separators. 
We compare 1) \textbf{Prompt}, which uses a textual indicator for each unimodal feature (see Equation~\ref{eq:textual_prompt}), 
and 2) \textbf{Fusion}, which utilizes a self-attention mechanism to obtain a fused token (see Equation~\ref{eq:fuse_modal}) \textit{w.r.t.} HitRate@1.
To distinguish the textual tokens and multimodal tokens, we include different separation strategies for <Sep>:
\begin{enumerate}
    \item \textbf{Textual Sep.}: Use a textual indicator (\eg <Sep> := "content token: ").
    \item \textbf{No Sep.}: Directly concatenate these two types of tokens.
    \item \textbf{Soft Sep.}: Train an embedding dynamically adapted for different contexts.
\end{enumerate}

Table~\ref{tab:process_and_sep} illustrates the detailed results. 
It is clear that the direct prompt-based token organization is inferior to the nuanced fusion method, \ie \textbf{Fusion} > \textbf{Prompt},  which contributes to finding the attentive features that adapt to varied bundle strategies.
There are also alternative methods to complete the hybrid tokens of both textual description tokens and multimodal feature tokens. 
As expected, the simple textual separator is inferior to the other two approaches. 
However, we note that the benefit of \textbf{Soft Sep.} is lower than \textbf{No Sep.} in the POG and POG\_dense datasets. 
This can be attributed to the well-categorized descriptions of items being consistent with the semantic multimodal tokens, where any extra separation might break such consistency. 
In contrast, for the discrepancy in Spotify and Spotify\_sparse, a dynamic soft separator outperforms.

\begin{table}
\renewcommand{\arraystretch}{1.1}
\setlength{\tabcolsep}{1mm}
\centering
\caption{The performance comparison $\textit{w.r.t}$ different multimodal token processes and seperators.}
\vspace{-2mm}
\label{tab:process_and_sep}
\begin{tabular}{l|c|c|c|c} 
\toprule
\textbf{Setting}      & \textbf{POG}                               & \textbf{POG\_dense}                        & \textbf{Spotify}         & \textbf{Spotify\_sparse}  \\ 
\midrule
\multicolumn{5}{l}{\textit{Different multimodal token processes}}                                                                                                      \\ 
\midrule
\textbf{Prompt}       & 0.2799                                     & 0.6133                                    & 0.8076                  & 0.7731                   \\
\textbf{Fusion}       & \textbf{0.4131}                            & \textbf{0.6675}                            & \textbf{0.8242}          & \textbf{0.7959}           \\ 
\midrule
\multicolumn{5}{l}{\textit{Different seperators}}                                                                                                                      \\ 
\midrule
\textbf{Textual Sep.} & 0.3631                                     & 0.6243                                     & 0.7528                   & 0.7489                    \\
\textbf{No Sep.}      & \textbf{\textbf{\textbf{\textbf{0.4131}}}} & \textbf{\textbf{\textbf{\textbf{0.6675}}}} & 0.8183                   & 0.7778                    \\
\textbf{Soft Sep.}    & 0.4077                                     & 0.6460                                     & \textbf{\textbf{0.8242}} & \textbf{\textbf{0.7959}}  \\
\bottomrule
\end{tabular}
\vspace{-2mm}
\end{table}

\subsubsection{The impact of progressive optimization}

\begin{figure}
    \centering
    \includegraphics[width=\linewidth]{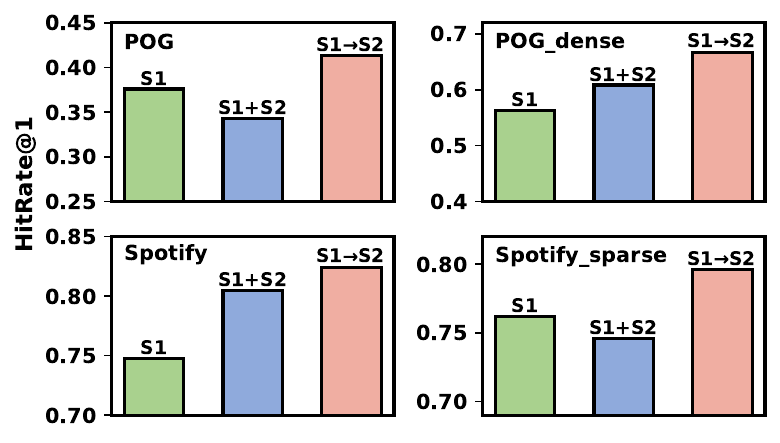}
    \vspace{-6mm}
    \caption{The performance \textit{w.r.t.} different optimization stages.}
    \label{fig:stages}
    \vspace{-4mm}
\end{figure}

To emphasize the necessity and effectiveness of disentangled optimization, we conducted experiments on different optimization stages:
\begin{enumerate}
    \item \textbf{S1}: Optimization stage for bundle pattern activation (see Equation~\ref{eq:s1}).
    \item \textbf{S2}: Optimization stage for multimodal semantic understanding (see Equation~\ref{eq:s2}).
    \item \textbf{S1+S2}: Directly combining both objectives and optimizing $\Theta_{\text{LoRA}}$ and $\Theta_{\text{F}}$ simultaneously.
    \item \textbf{S1$\to$S2}: Utilizing the progressive strategy to optimize these two disentangled objectives, adhering to our proposed progressive optimization in Section ~\ref{sec:optimization}.
\end{enumerate}
Since the single \textbf{S2} does not optimize the product bundling capability of LLMs, showing low instruction-following performance, we exclude \textbf{S2} from comparison in Figure~\ref{fig:stages}. 

Overall, \textbf{S1$\to$S2} outperforms all other optimization strategies, which can be attributed to progressively optimizing disentangled objectives, facilitating the successful adaptation of LLMs to multimodal product bundling. 
Notably, \textbf{S1+S2} outperforms \textbf{S1} when denser relational data are provided in POG\_dense and Spotify compared to their sparse versions, POG and Spotify\_sparse, which show inconsistent optimization that leads to worse performance. 
This finding further demonstrates that our optimization method not only alleviates the optimizing inconsistency in sparse datasets but also enhances performance improvements in dense datasets.

\subsection{Model Study (RQ3)}
For a deeper understanding of our proposed methods and to further address RQ3, we conduct experiments under various settings, particularly focusing on different candidate sizes and the cold-start scenario. 
These experiments explore the capabilities of our methods in handling varied item lengths and invoking extensive knowledge for cold product bundling. 
Moreover, we analyze progressive optimization for improving effectiveness during training and our proposed multimodal token process for boosting efficiency during inference. 
Finally, we demonstrate comparison with existing MLLMs and the Scalability of Bundle-MLLM.

\subsubsection{The Impact of candidate size}
\label{sec:candidates}

\begin{figure}
    \centering
    \includegraphics[width=\linewidth]{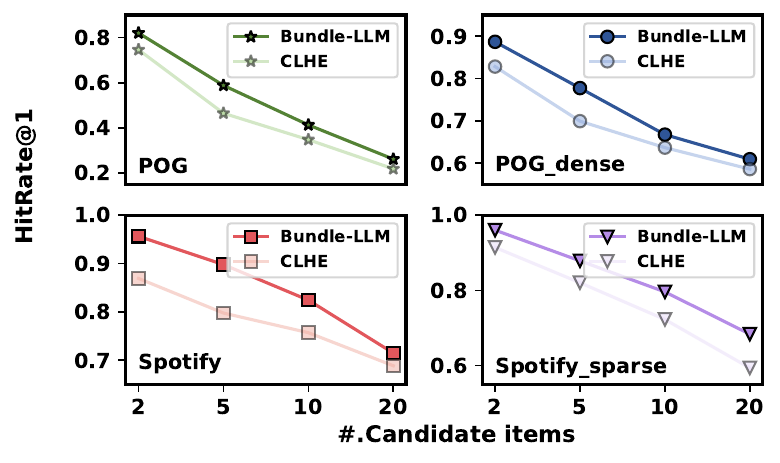}
    \vspace{-6mm}
    \caption{The performance \textit{w.r.t.} different numbers of candidate items.}
    \label{fig:candidates}
\end{figure}

We conduct experiments with different numbers of candidate items for Bundle-MLLM, as shown in Figure~\ref{fig:candidates}. 
As the number of candidate items increases, the task's difficulty rises and the prediction accuracy decreases. Nonetheless, Bundle-MLLM consistently outperforms the SOTA method (\ie CLHE) in all cases. 
This demonstrates the stability of our proposed method and highlights its potential to handle a large number of candidate items effectively.

\subsubsection{Cold-setting performance}
Even with a few samples, Bundle-MLLM achieves superior performance, prompting an exploration of the cold-setting scenario through a comparison with CLHE. In this analysis, we re-split the POG dataset to ensure no overlap between training and testing items.
As illustrated in Table~\ref{tab:cold}, compared to CLHE, the improvement is significant and increases with the number of candidate items. 
Due to its reliance on sufficient supervised learning signals, CLHE shows a distinct gap in performance when evaluated on a testing set that is significantly out-of-distribution from the training set, leading to inferior results.
In contrast, Bundle-MLLM, with its extensive knowledge and activated bundle pattern, can still effectively tackle the product bundling task in a cold-setting scenario.

\begin{table}
\centering
\caption{The performance (HitRate@1) compared with CLHE in the cold-setting.}
\setlength{\tabcolsep}{2.5mm}
\label{tab:cold}
\begin{tabular}{l|cccc} 
\toprule
\textbf{\#Candidates}    & \textbf{2}       & \textbf{5}       & \textbf{10}      & \textbf{20}       \\ 
\midrule
\textbf{CLHE}       & 0.6398          & 0.2617          & 0.1061          & 0.0239           \\ 
\textbf{Bundle-MLLM} & \textbf{0.9414} & \textbf{0.8047} & \textbf{0.6172} & \textbf{0.4688}  \\ 
\midrule
\textbf{Rel.Imp.}   & 47.14\%          & 207.49\%         & 481.72\%         & 1861.51\%         \\
\bottomrule
\end{tabular}
\vspace{-2mm}
\end{table}

\subsubsection{Training procedure}

\begin{figure}
    \centering
    \includegraphics[width=\linewidth]{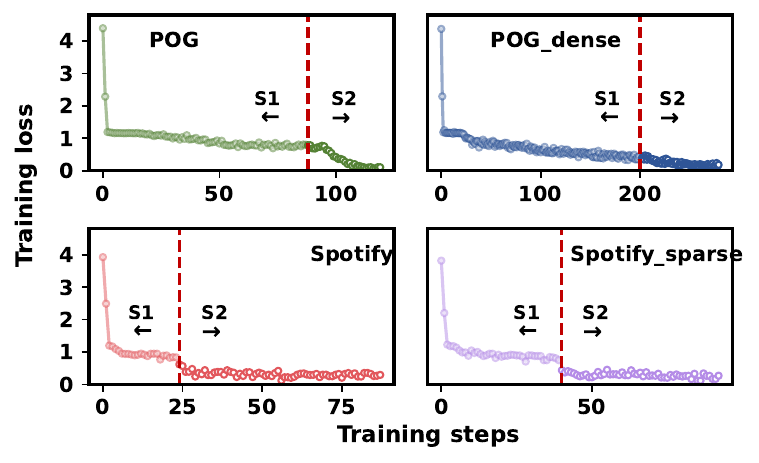}
    \vspace{-8mm}
    \caption{The training loss curve across different optimization stages.}
    \vspace{-4mm}
    \label{fig:training_loss}
\end{figure}

To elucidate the progressive optimization process, we illustrate the training loss curve across different stages on four datasets in Figure~\ref{fig:training_loss}. 
Our progressive optimization strategy first optimizes the LoRA using only textual information (\ie \textbf{S1}), followed by optimizing the multimodal token process module (\ie \textbf{S2}). 
This endows Bundle-MLLM with multimodal semantic understanding, further enhancing its product bundling capability.
From this, we observe the gradual smoothing of the curves during each stage and a significant drop in loss when transitioning to the next optimization stage. 
This transition breaks the standstill of the previous status, further aligning Bundle-MLLM to product bundling tasks. 
These observations firmly demonstrate the rationality and effectiveness of our proposed optimization strategy.

\subsubsection{Inference efficiency}
\begin{figure}
    \centering
    \includegraphics[width=\linewidth]{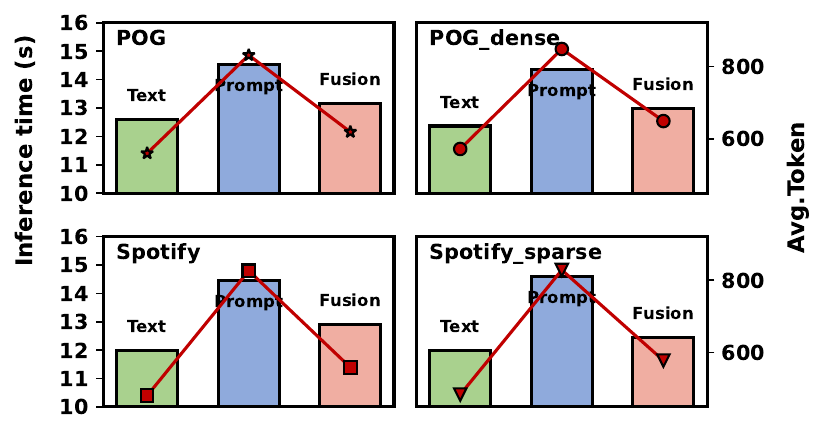}
    \vspace{-6mm}
    \caption{Inference time (bar) and average number of tokens (line) \emph{w.r.t.} different item tokenization strategies.}
    \vspace{-4mm}
    \label{fig:efficiency}
\end{figure}

Benefiting from our proposed multimodal token process, we improve the efficiency of autoregressive inference in Bundle-MLLM. 
For a comprehensive comparison, we calculate the inference time and the average number of tokens for each bundling query using \textbf{Text} (only text information provided), \textbf{Prompt}, and \textbf{Fusion} (see Section ~\ref{sec:fusion_seperator}), as shown in Figure~\ref{fig:efficiency}.
Compared to \textbf{Text}, there is a slight increase in time and token count for \textbf{Fusion} due to the incorporation of multiple data formats. 
However, when compared to \textbf{Prompt}, which provides the same information, \textbf{Fusion} significantly reduces both the time and token cost. 
This efficiency gain becomes more pronounced as the number of items in both the input partial bundle and candidate sets increases.

\subsubsection{Comparison with existing MLLMs} we include comparisons with three well-recognized MLLMs: MiniCPM, QwenVL, and Llava, using the POG dataset.
Our comparison highlights that while these MLLMs can improve performance when vision support is incorporated, they still fall short in handling the bundling task effectively. As shown in Table~\ref{tab:mllms_comparison}, Bundle-MLLM significantly outperforms recent pre-trained MLLMs, achieving a HitRate@1 of 0.4131 compared to 0.1689 for the closest competitor.

\begin{table}
\centering
\caption{The performance comparison with existing MLLMs on POG dataset.}
\vspace{-3mm}
\label{tab:mllms_comparison}
\begin{tabular}{l|cccc} 
\toprule
\textbf{Model} & \textbf{Bundle-MLLM} & \textbf{MiniCPM} & \textbf{QwenVL} & \textbf{Llava}  \\ 
\midrule
HitRate@1      & \textbf{0.4131}      & 0.1689           & 0.1216          & 0.0952          \\
\bottomrule
\end{tabular}
\vspace{-2mm}
\end{table}

\subsubsection{Scalability} Bundle-MLLM shows the potential scalebility, demonstrated by the performance improvement when replacing the base LLM from Llama2-7B to Llama2-13B, as shown in Table~\ref{tab:base_model}.
This indicates that Bundle-MLLM can boost a larger LLM for bundling by leveraging its extensive knowledge and stronger reasoning capability.
However, low parameters will constraint the bundling performance (\eg, 0.1113 for 2B model). This is reasonable that we use few-shot tuning for efficiency, while 2B model has inherently limited knowledge that can not handle bundling tasks. 

\begin{table}
\centering
\caption{The performance comparison on using different LLM backbones for Bundle-MLLM.}
\vspace{-3mm}
\label{tab:base_model}
\begin{tabular}{l|ccc} 
\toprule
\textbf{LLM Backbone} & \textbf{Gemma-2B} & \textbf{Llama2-7B} & \textbf{Llama2-13B}  \\ 
\midrule
HitRate@1 & 0.1113            & 0.4131             & \textbf{0.4492}      \\
\bottomrule
\end{tabular}
\vspace{-2mm}
\end{table}

\section{Related work}
We review the literature on product bundling and multimodal large language models.

\subsection{Bundle Recommendation \& Construction}

The development of product bundling has evolved into two main categories: recommending pre-built bundles~\cite{BundleNet,HFGN} (\ie bundle recommendation) and generating bundles~\cite{BYOB, BGN} (\ie bundle construction/product bundling)~\cite{revisit}. 
Initially, bundle recommendation focused on transferring benefits from item recommendation~\cite{Lightgcn, EliMRec, Liu2024PreferDiff} to bundle recommendation~\cite{HFGN}. 
Recent advancements include contrastive-based frameworks~\cite{CrossCBR,HIDGN, MultiCBR}, deep attentive multi-task models, transformer-based models for bundle recommendation~\cite{BundleGT} and distillation-based method~\cite{DGMAE}.
Bundle construction aims to complete the given partial bundles, \ie seed items, and retrieve the remaining items from the candidates~\cite{BGN, CLHE}.
Concerning different intents, it varies across different domains like fashion outfit~\cite{FARM}, music playlist~\cite{music_continuation, LARP}, and etc~\cite{MealRec}. 
To achieve dynamical bundle construction, user behaviors are incorporate for personalization and offer informative bundle patterns hidden in user-item interactions~\cite{BYOB,List_continuation, BGGN}, 
or integrate visual information to bundle complementary items~\cite{POG}.
And most recent method leverages both user feedback and mutlimodal features to aid the insufficiency bundle-item affiliations and cold-item issues~\cite{CLHE}. 
There also a single paper that utilizes LLMs with in-context learning via item titles and neighbor sessions~\cite{bundleICL}. 
Different from these methods, we exhibit the semantics from multiple data formats to embody the bundle strategies, paired with activated LLMs, thus improving the bundle construction capability. 

\subsection{Multimodal Large Language Model}

Language models (LMs) have seen a significant transformation, particularly with the development of large language models (LLMs) that harness advanced neural network architectures like transformers~\cite{transformer} and pre-training on expansive datasets.
Models like GPT~\cite{GPT} leverage these technologies to grasp nuanced language patterns and generate responses with a deep contextual awareness, empowering various domains~\cite{Liu2025LRCD,AlphaRec, SDPO}.
Building on the success of LLMs and PEFT~\cite{LoRA} to facilitate the training efficiency, the field has expanded into the realm of multimodal large language models (MLLMs)~\cite{surveyMLLM}. 
Since LLMs can only perceive text, bridging the gap between natural language and other modalities falls into two categories:
1) utilizing a learnable interface to project information into the space that LLM can understand~\cite{minigpt4, llava}, 
and 2) leveraging the textual proxy~\cite{zhu2023chatgpt} through expert models, such as image captioning model~\cite{blip2}.
These models integrate different formats of data, like vision~\cite{minigpt4}, audio~\cite{Macaw-llm} and graph~\cite{KG_LLM, gnn_Adapter, GraphTranslator} into an unified framework.
Inspired by its complex reasoning on different modalities, MLLMs expand to diverse domains like biomedicine~\cite{Llava-med}, robotics~\cite{Palm-e} and recommendation system~\cite{tallrec, Llara}.
In our work, we further step in leveraging multimodal language model techniques to bundle construction, focusing on activating untapped bundle pattern with multimodal semantics via progressive optimization.

\section{Conclusion and Future work}
In this work, we pioneered a novel multimodal large language framework for product bundling, termed Bundle-MLLM, which integrates multiple data formats (\ie textual, visual/acoustic, and relational information).
Notably, Bundle-MLLM surpasses vanilla LLM adaptations by activating untapped bundle patterns and compensating for sole textual effects.
Bundle-MLLM employs advanced encoders to extract heterogeneous features and utilizes a simple yet effective multimodal process module to integrate them into a single token, which is then organized within a hybrid item tokenization scheme. 
By framing the product bundling task as a multiple-choice question, we implemented progressive optimization to achieve effective product bundling capability with comprehensive multimodal semantic understanding of LLMs.
Empirical results demonstrated that Bundle-MLLM significantly outperforms all baselines in product bundling, underscoring its effectiveness and superior performance. 
Ablation studies and model analyses further validated the rationale and efficacy of each design component.
This work marks an initial step in endowing LLMs with multimodal semantic understanding in product bundling. 
Moving forward, we will steer the future focus to 1) efficiently tackling much longer candidates and 2) endowing LLMs with personalized bundling ability. 
And we hope that Bundle-MLLM can inspire the researchers to explore a more unified framework, paving the way to the development of versatile product bundling.

\section*{Acknowledgment}
This research is supported by the Fundamental Research Funds for the Central Universities (No. CUC24QT19).
This research is also
supported by NExT Research
Center.

\bibliographystyle{ACM-Reference-Format}
\balance
\bibliography{sample-base}


\begin{thebibliography}{54}


\ifx \showCODEN    \undefined \def \showCODEN     #1{\unskip}     \fi
\ifx \showDOI      \undefined \def \showDOI       #1{#1}\fi
\ifx \showISBNx    \undefined \def \showISBNx     #1{\unskip}     \fi
\ifx \showISBNxiii \undefined \def \showISBNxiii  #1{\unskip}     \fi
\ifx \showISSN     \undefined \def \showISSN      #1{\unskip}     \fi
\ifx \showLCCN     \undefined \def \showLCCN      #1{\unskip}     \fi
\ifx \shownote     \undefined \def \shownote      #1{#1}          \fi
\ifx \showarticletitle \undefined \def \showarticletitle #1{#1}   \fi
\ifx \showURL      \undefined \def \showURL       {\relax}        \fi
\providecommand\bibfield[2]{#2}
\providecommand\bibinfo[2]{#2}
\providecommand\natexlab[1]{#1}
\providecommand\showeprint[2][]{arXiv:#2}

\bibitem[Achiam et~al\mbox{.}(2023)]%
        {gpt4}
\bibfield{author}{\bibinfo{person}{Josh Achiam}, \bibinfo{person}{Steven Adler}, \bibinfo{person}{Sandhini Agarwal}, \bibinfo{person}{Lama Ahmad}, \bibinfo{person}{Ilge Akkaya}, \bibinfo{person}{Florencia~Leoni Aleman}, \bibinfo{person}{Diogo Almeida}, \bibinfo{person}{Janko Altenschmidt}, \bibinfo{person}{Sam Altman}, \bibinfo{person}{Shyamal Anadkat}, {et~al\mbox{.}}} \bibinfo{year}{2023}\natexlab{}.
\newblock \showarticletitle{Gpt-4 technical report}.
\newblock \bibinfo{journal}{\emph{arXiv preprint arXiv:2303.08774}} (\bibinfo{year}{2023}).
\newblock


\bibitem[Bai et~al\mbox{.}(2019)]%
        {BGN}
\bibfield{author}{\bibinfo{person}{Jinze Bai}, \bibinfo{person}{Chang Zhou}, \bibinfo{person}{Junshuai Song}, \bibinfo{person}{Xiaoru Qu}, \bibinfo{person}{Weiting An}, \bibinfo{person}{Zhao Li}, {and} \bibinfo{person}{Jun Gao}.} \bibinfo{year}{2019}\natexlab{}.
\newblock \showarticletitle{Personalized bundle list recommendation}. In \bibinfo{booktitle}{\emph{{WWW}}}. \bibinfo{pages}{60--71}.
\newblock


\bibitem[Bakos and Brynjolfsson(2000)]%
        {bakos2000bundling}
\bibfield{author}{\bibinfo{person}{Yannis Bakos} {and} \bibinfo{person}{Erik Brynjolfsson}.} \bibinfo{year}{2000}\natexlab{}.
\newblock \showarticletitle{Bundling and Competition on the Internet}.
\newblock \bibinfo{journal}{\emph{Marketing science}} \bibinfo{volume}{19}, \bibinfo{number}{1} (\bibinfo{year}{2000}), \bibinfo{pages}{63--82}.
\newblock


\bibitem[Bao et~al\mbox{.}(2023)]%
        {tallrec}
\bibfield{author}{\bibinfo{person}{Keqin Bao}, \bibinfo{person}{Jizhi Zhang}, \bibinfo{person}{Yang Zhang}, \bibinfo{person}{Wenjie Wang}, \bibinfo{person}{Fuli Feng}, {and} \bibinfo{person}{Xiangnan He}.} \bibinfo{year}{2023}\natexlab{}.
\newblock \showarticletitle{Tallrec: An effective and efficient tuning framework to align large language model with recommendation}. In \bibinfo{booktitle}{\emph{{RecSys}}}. \bibinfo{pages}{1007--1014}.
\newblock


\bibitem[Chang et~al\mbox{.}(2021)]%
        {BGGN}
\bibfield{author}{\bibinfo{person}{Jianxin Chang}, \bibinfo{person}{Chen Gao}, \bibinfo{person}{Xiangnan He}, \bibinfo{person}{Depeng Jin}, {and} \bibinfo{person}{Yong Li}.} \bibinfo{year}{2021}\natexlab{}.
\newblock \showarticletitle{Bundle recommendation and generation with graph neural networks}.
\newblock \bibinfo{journal}{\emph{{TKDE}}} \bibinfo{volume}{35}, \bibinfo{number}{3} (\bibinfo{year}{2021}), \bibinfo{pages}{2326--2340}.
\newblock


\bibitem[Chen et~al\mbox{.}(2019)]%
        {POG}
\bibfield{author}{\bibinfo{person}{Wen Chen}, \bibinfo{person}{Pipei Huang}, \bibinfo{person}{Jiaming Xu}, \bibinfo{person}{Xin Guo}, \bibinfo{person}{Cheng Guo}, \bibinfo{person}{Fei Sun}, \bibinfo{person}{Chao Li}, \bibinfo{person}{Andreas Pfadler}, \bibinfo{person}{Huan Zhao}, {and} \bibinfo{person}{Binqiang Zhao}.} \bibinfo{year}{2019}\natexlab{}.
\newblock \showarticletitle{POG: personalized outfit generation for fashion recommendation at Alibaba iFashion}. In \bibinfo{booktitle}{\emph{{KDD}}}. \bibinfo{pages}{2662--2670}.
\newblock


\bibitem[Chu et~al\mbox{.}(2023)]%
        {qwen_audio}
\bibfield{author}{\bibinfo{person}{Yunfei Chu}, \bibinfo{person}{Jin Xu}, \bibinfo{person}{Xiaohuan Zhou}, \bibinfo{person}{Qian Yang}, \bibinfo{person}{Shiliang Zhang}, \bibinfo{person}{Zhijie Yan}, \bibinfo{person}{Chang Zhou}, {and} \bibinfo{person}{Jingren Zhou}.} \bibinfo{year}{2023}\natexlab{}.
\newblock \showarticletitle{Qwen-Audio: Advancing Universal Audio Understanding via Unified Large-Scale Audio-Language Models}.
\newblock \bibinfo{journal}{\emph{CoRR}}  \bibinfo{volume}{abs/2311.07919} (\bibinfo{year}{2023}).
\newblock
\showeprint[arXiv]{2311.07919}


\bibitem[Deng et~al\mbox{.}(2021)]%
        {BYOB}
\bibfield{author}{\bibinfo{person}{Qilin Deng}, \bibinfo{person}{Kai Wang}, \bibinfo{person}{Minghao Zhao}, \bibinfo{person}{Runze Wu}, \bibinfo{person}{Yu Ding}, \bibinfo{person}{Zhene Zou}, \bibinfo{person}{Yue Shang}, \bibinfo{person}{Jianrong Tao}, {and} \bibinfo{person}{Changjie Fan}.} \bibinfo{year}{2021}\natexlab{}.
\newblock \showarticletitle{Build your own bundle-a neural combinatorial optimization method}. In \bibinfo{booktitle}{\emph{{ACM} Multimedia}}. \bibinfo{pages}{2625--2633}.
\newblock


\bibitem[Deng et~al\mbox{.}(2020)]%
        {BundleNet}
\bibfield{author}{\bibinfo{person}{Qilin Deng}, \bibinfo{person}{Kai Wang}, \bibinfo{person}{Minghao Zhao}, \bibinfo{person}{Zhene Zou}, \bibinfo{person}{Runze Wu}, \bibinfo{person}{Jianrong Tao}, \bibinfo{person}{Changjie Fan}, {and} \bibinfo{person}{Liang Chen}.} \bibinfo{year}{2020}\natexlab{}.
\newblock \showarticletitle{Personalized bundle recommendation in online games}. In \bibinfo{booktitle}{\emph{{CIKM}}}. \bibinfo{pages}{2381--2388}.
\newblock


\bibitem[Driess et~al\mbox{.}(2023)]%
        {Palm-e}
\bibfield{author}{\bibinfo{person}{Danny Driess}, \bibinfo{person}{Fei Xia}, \bibinfo{person}{Mehdi~SM Sajjadi}, \bibinfo{person}{Corey Lynch}, \bibinfo{person}{Aakanksha Chowdhery}, \bibinfo{person}{Brian Ichter}, \bibinfo{person}{Ayzaan Wahid}, \bibinfo{person}{Jonathan Tompson}, \bibinfo{person}{Quan Vuong}, \bibinfo{person}{Tianhe Yu}, {et~al\mbox{.}}} \bibinfo{year}{2023}\natexlab{}.
\newblock \showarticletitle{Palm-e: An embodied multimodal language model}.
\newblock \bibinfo{journal}{\emph{arXiv preprint arXiv:2303.03378}} (\bibinfo{year}{2023}).
\newblock


\bibitem[Elizalde et~al\mbox{.}(2023)]%
        {clap}
\bibfield{author}{\bibinfo{person}{Benjamin Elizalde}, \bibinfo{person}{Soham Deshmukh}, \bibinfo{person}{Mahmoud Al~Ismail}, {and} \bibinfo{person}{Huaming Wang}.} \bibinfo{year}{2023}\natexlab{}.
\newblock \showarticletitle{Clap learning audio concepts from natural language supervision}. In \bibinfo{booktitle}{\emph{{ICASSP}}}. IEEE, \bibinfo{pages}{1--5}.
\newblock


\bibitem[Han et~al\mbox{.}(2017)]%
        {BiLSTM}
\bibfield{author}{\bibinfo{person}{Xintong Han}, \bibinfo{person}{Zuxuan Wu}, \bibinfo{person}{Yu-Gang Jiang}, {and} \bibinfo{person}{Larry~S Davis}.} \bibinfo{year}{2017}\natexlab{}.
\newblock \showarticletitle{Learning fashion compatibility with bidirectional lstms}. In \bibinfo{booktitle}{\emph{{ACM} Multimedia}}. \bibinfo{pages}{1078--1086}.
\newblock


\bibitem[He et~al\mbox{.}(2020a)]%
        {Lightgcn}
\bibfield{author}{\bibinfo{person}{Xiangnan He}, \bibinfo{person}{Kuan Deng}, \bibinfo{person}{Xiang Wang}, \bibinfo{person}{Yan Li}, \bibinfo{person}{Yongdong Zhang}, {and} \bibinfo{person}{Meng Wang}.} \bibinfo{year}{2020}\natexlab{a}.
\newblock \showarticletitle{Lightgcn: Simplifying and powering graph convolution network for recommendation}. In \bibinfo{booktitle}{\emph{{SIGIR}}}. \bibinfo{pages}{639--648}.
\newblock


\bibitem[He et~al\mbox{.}(2020b)]%
        {List_continuation}
\bibfield{author}{\bibinfo{person}{Yun He}, \bibinfo{person}{Yin Zhang}, \bibinfo{person}{Weiwen Liu}, {and} \bibinfo{person}{James Caverlee}.} \bibinfo{year}{2020}\natexlab{b}.
\newblock \showarticletitle{Consistency-aware recommendation for user-generated item list continuation}. In \bibinfo{booktitle}{\emph{{WSDM}}}. \bibinfo{pages}{250--258}.
\newblock


\bibitem[Hu et~al\mbox{.}(2021)]%
        {LoRA}
\bibfield{author}{\bibinfo{person}{Edward~J Hu}, \bibinfo{person}{Yelong Shen}, \bibinfo{person}{Phillip Wallis}, \bibinfo{person}{Zeyuan Allen-Zhu}, \bibinfo{person}{Yuanzhi Li}, \bibinfo{person}{Shean Wang}, \bibinfo{person}{Lu Wang}, {and} \bibinfo{person}{Weizhu Chen}.} \bibinfo{year}{2021}\natexlab{}.
\newblock \showarticletitle{Lora: Low-rank adaptation of large language models}.
\newblock \bibinfo{journal}{\emph{arXiv preprint arXiv:2106.09685}} (\bibinfo{year}{2021}).
\newblock


\bibitem[Huang et~al\mbox{.}(2024)]%
        {gnn_Adapter}
\bibfield{author}{\bibinfo{person}{Xuanwen Huang}, \bibinfo{person}{Kaiqiao Han}, \bibinfo{person}{Yang Yang}, \bibinfo{person}{Dezheng Bao}, \bibinfo{person}{Quanjin Tao}, \bibinfo{person}{Ziwei Chai}, {and} \bibinfo{person}{Qi Zhu}.} \bibinfo{year}{2024}\natexlab{}.
\newblock \showarticletitle{Can GNN be Good Adapter for LLMs?}
\newblock  (\bibinfo{year}{2024}).
\newblock


\bibitem[Lester et~al\mbox{.}(2021)]%
        {prompt-tuning}
\bibfield{author}{\bibinfo{person}{Brian Lester}, \bibinfo{person}{Rami Al-Rfou}, {and} \bibinfo{person}{Noah Constant}.} \bibinfo{year}{2021}\natexlab{}.
\newblock \showarticletitle{The Power of Scale for Parameter-Efficient Prompt Tuning}. In \bibinfo{booktitle}{\emph{{EMNLP}}}. \bibinfo{pages}{3045--3059}.
\newblock


\bibitem[Li et~al\mbox{.}(2024b)]%
        {Llava-med}
\bibfield{author}{\bibinfo{person}{Chunyuan Li}, \bibinfo{person}{Cliff Wong}, \bibinfo{person}{Sheng Zhang}, \bibinfo{person}{Naoto Usuyama}, \bibinfo{person}{Haotian Liu}, \bibinfo{person}{Jianwei Yang}, \bibinfo{person}{Tristan Naumann}, \bibinfo{person}{Hoifung Poon}, {and} \bibinfo{person}{Jianfeng Gao}.} \bibinfo{year}{2024}\natexlab{b}.
\newblock \showarticletitle{Llava-med: Training a large language-and-vision assistant for biomedicine in one day}.
\newblock \bibinfo{journal}{\emph{{NeurIPS}}} (\bibinfo{year}{2024}).
\newblock


\bibitem[Li et~al\mbox{.}(2023)]%
        {blip2}
\bibfield{author}{\bibinfo{person}{Junnan Li}, \bibinfo{person}{Dongxu Li}, \bibinfo{person}{Silvio Savarese}, {and} \bibinfo{person}{Steven Hoi}.} \bibinfo{year}{2023}\natexlab{}.
\newblock \showarticletitle{Blip-2: Bootstrapping language-image pre-training with frozen image encoders and large language models}. In \bibinfo{booktitle}{\emph{{ICML}}}. \bibinfo{pages}{19730--19742}.
\newblock


\bibitem[Li et~al\mbox{.}(2024a)]%
        {MealRec}
\bibfield{author}{\bibinfo{person}{Ming Li}, \bibinfo{person}{Lin Li}, \bibinfo{person}{Xiaohui Tao}, {and} \bibinfo{person}{Jimmy~Xiangji Huang}.} \bibinfo{year}{2024}\natexlab{a}.
\newblock \showarticletitle{MealRec: A Meal Recommendation Dataset with Meal-Course Affiliation for Personalization and Healthiness}.
\newblock  (\bibinfo{year}{2024}).
\newblock


\bibitem[Li et~al\mbox{.}(2020)]%
        {HFGN}
\bibfield{author}{\bibinfo{person}{Xingchen Li}, \bibinfo{person}{Xiang Wang}, \bibinfo{person}{Xiangnan He}, \bibinfo{person}{Long Chen}, \bibinfo{person}{Jun Xiao}, {and} \bibinfo{person}{Tat-Seng Chua}.} \bibinfo{year}{2020}\natexlab{}.
\newblock \showarticletitle{Hierarchical fashion graph network for personalized outfit recommendation}. In \bibinfo{booktitle}{\emph{{SIGIR}}}. \bibinfo{pages}{159--168}.
\newblock


\bibitem[Liang et~al\mbox{.}(2018)]%
        {multiVAE}
\bibfield{author}{\bibinfo{person}{Dawen Liang}, \bibinfo{person}{Rahul~G Krishnan}, \bibinfo{person}{Matthew~D Hoffman}, {and} \bibinfo{person}{Tony Jebara}.} \bibinfo{year}{2018}\natexlab{}.
\newblock \showarticletitle{Variational autoencoders for collaborative filtering}. In \bibinfo{booktitle}{\emph{{WWW}}}. \bibinfo{pages}{689--698}.
\newblock


\bibitem[Liao et~al\mbox{.}(2024)]%
        {Llara}
\bibfield{author}{\bibinfo{person}{Jiayi Liao}, \bibinfo{person}{Sihang Li}, \bibinfo{person}{Zhengyi Yang}, \bibinfo{person}{Jiancan Wu}, \bibinfo{person}{Yancheng Yuan}, \bibinfo{person}{Xiang Wang}, {and} \bibinfo{person}{Xiangnan He}.} \bibinfo{year}{2024}\natexlab{}.
\newblock \showarticletitle{Llara: Aligning large language models with sequential recommenders}.
\newblock  (\bibinfo{year}{2024}).
\newblock


\bibitem[Lin et~al\mbox{.}(2019)]%
        {FARM}
\bibfield{author}{\bibinfo{person}{Yujie Lin}, \bibinfo{person}{Pengjie Ren}, \bibinfo{person}{Zhumin Chen}, \bibinfo{person}{Zhaochun Ren}, \bibinfo{person}{Jun Ma}, {and} \bibinfo{person}{Maarten De~Rijke}.} \bibinfo{year}{2019}\natexlab{}.
\newblock \showarticletitle{Improving outfit recommendation with co-supervision of fashion generation}. In \bibinfo{booktitle}{\emph{{WWW}}}. \bibinfo{pages}{1095--1105}.
\newblock


\bibitem[Liu et~al\mbox{.}(2023)]%
        {llava}
\bibfield{author}{\bibinfo{person}{Haotian Liu}, \bibinfo{person}{Chunyuan Li}, \bibinfo{person}{Qingyang Wu}, {and} \bibinfo{person}{Yong~Jae Lee}.} \bibinfo{year}{2023}\natexlab{}.
\newblock \showarticletitle{Visual Instruction Tuning}. In \bibinfo{booktitle}{\emph{NeurIPS}}.
\newblock


\bibitem[Liu et~al\mbox{.}(2024)]%
        {Liu2024PreferDiff}
\bibfield{author}{\bibinfo{person}{Shuo Liu}, \bibinfo{person}{An Zhang}, \bibinfo{person}{Guoqing Hu}, \bibinfo{person}{Hong Qian}, {and} \bibinfo{person}{Tat-seng Chua}.} \bibinfo{year}{2024}\natexlab{}.
\newblock \showarticletitle{Preference Diffusion for Recommendation}.
\newblock \bibinfo{journal}{\emph{arXiv preprint arXiv:2410.13117}} (\bibinfo{year}{2024}).
\newblock


\bibitem[Liu et~al\mbox{.}(2025)]%
        {Liu2025LRCD}
\bibfield{author}{\bibinfo{person}{Shuo Liu}, \bibinfo{person}{Zihan Zhou}, \bibinfo{person}{Yuanhao Liu}, \bibinfo{person}{Jing Zhang}, {and} \bibinfo{person}{Hong Qian}.} \bibinfo{year}{2025}\natexlab{}.
\newblock \showarticletitle{Language Representation Favored Zero-Shot Cross-Domain Cognitive Diagnosis}. In \bibinfo{booktitle}{\emph{KDD}}.
\newblock


\bibitem[Liu et~al\mbox{.}(2022)]%
        {EliMRec}
\bibfield{author}{\bibinfo{person}{Xiaohao Liu}, \bibinfo{person}{Zhulin Tao}, \bibinfo{person}{Jiahong Shao}, \bibinfo{person}{Lifang Yang}, {and} \bibinfo{person}{Xianglin Huang}.} \bibinfo{year}{2022}\natexlab{}.
\newblock \showarticletitle{Elimrec: Eliminating single-modal bias in multimedia recommendation}. In \bibinfo{booktitle}{\emph{{ACM} Multimedia}}. \bibinfo{pages}{687--695}.
\newblock


\bibitem[Lyu et~al\mbox{.}(2023)]%
        {Macaw-llm}
\bibfield{author}{\bibinfo{person}{Chenyang Lyu}, \bibinfo{person}{Minghao Wu}, \bibinfo{person}{Longyue Wang}, \bibinfo{person}{Xinting Huang}, \bibinfo{person}{Bingshuai Liu}, \bibinfo{person}{Zefeng Du}, \bibinfo{person}{Shuming Shi}, {and} \bibinfo{person}{Zhaopeng Tu}.} \bibinfo{year}{2023}\natexlab{}.
\newblock \showarticletitle{Macaw-llm: Multi-modal language modeling with image, audio, video, and text integration}.
\newblock \bibinfo{journal}{\emph{arXiv preprint arXiv:2306.09093}} (\bibinfo{year}{2023}).
\newblock


\bibitem[Ma et~al\mbox{.}(2024a)]%
        {MultiCBR}
\bibfield{author}{\bibinfo{person}{Yunshan Ma}, \bibinfo{person}{Yingzhi He}, \bibinfo{person}{Xiang Wang}, \bibinfo{person}{Yinwei Wei}, \bibinfo{person}{Xiaoyu Du}, \bibinfo{person}{Yuyangzi Fu}, {and} \bibinfo{person}{Tat-Seng Chua}.} \bibinfo{year}{2024}\natexlab{a}.
\newblock \showarticletitle{MultiCBR: Multi-view Contrastive Learning for Bundle Recommendation}.
\newblock \bibinfo{journal}{\emph{ACM Transactions on Information Systems}} \bibinfo{volume}{42}, \bibinfo{number}{4} (\bibinfo{year}{2024}), \bibinfo{pages}{1--23}.
\newblock


\bibitem[Ma et~al\mbox{.}(2022)]%
        {CrossCBR}
\bibfield{author}{\bibinfo{person}{Yunshan Ma}, \bibinfo{person}{Yingzhi He}, \bibinfo{person}{An Zhang}, \bibinfo{person}{Xiang Wang}, {and} \bibinfo{person}{Tat-Seng Chua}.} \bibinfo{year}{2022}\natexlab{}.
\newblock \showarticletitle{CrossCBR: Cross-view contrastive learning for bundle recommendation}. In \bibinfo{booktitle}{\emph{{KDD}}}. \bibinfo{pages}{1233--1241}.
\newblock


\bibitem[Ma et~al\mbox{.}(2024b)]%
        {CLHE}
\bibfield{author}{\bibinfo{person}{Yunshan Ma}, \bibinfo{person}{Xiaohao Liu}, \bibinfo{person}{Yinwei Wei}, \bibinfo{person}{Zhulin Tao}, \bibinfo{person}{Xiang Wang}, {and} \bibinfo{person}{Tat-Seng Chua}.} \bibinfo{year}{2024}\natexlab{b}.
\newblock \showarticletitle{Leveraging multimodal features and item-level user feedback for bundle construction}. In \bibinfo{booktitle}{\emph{{WSDM}}}. \bibinfo{pages}{510--519}.
\newblock


\bibitem[Pan et~al\mbox{.}(2024)]%
        {KG_LLM}
\bibfield{author}{\bibinfo{person}{Shirui Pan}, \bibinfo{person}{Linhao Luo}, \bibinfo{person}{Yufei Wang}, \bibinfo{person}{Chen Chen}, \bibinfo{person}{Jiapu Wang}, {and} \bibinfo{person}{Xindong Wu}.} \bibinfo{year}{2024}\natexlab{}.
\newblock \showarticletitle{Unifying large language models and knowledge graphs: A roadmap}.
\newblock \bibinfo{journal}{\emph{IEEE Transactions on Knowledge and Data Engineering}} (\bibinfo{year}{2024}).
\newblock


\bibitem[Radford et~al\mbox{.}(2018)]%
        {GPT}
\bibfield{author}{\bibinfo{person}{Alec Radford}, \bibinfo{person}{Karthik Narasimhan}, \bibinfo{person}{Tim Salimans}, \bibinfo{person}{Ilya Sutskever}, {et~al\mbox{.}}} \bibinfo{year}{2018}\natexlab{}.
\newblock \showarticletitle{Improving language understanding by generative pre-training}.
\newblock  (\bibinfo{year}{2018}).
\newblock


\bibitem[Ren et~al\mbox{.}(2023)]%
        {DGMAE}
\bibfield{author}{\bibinfo{person}{Yuyang Ren}, \bibinfo{person}{Zhang Haonan}, \bibinfo{person}{Luoyi Fu}, \bibinfo{person}{Xinbing Wang}, {and} \bibinfo{person}{Chenghu Zhou}.} \bibinfo{year}{2023}\natexlab{}.
\newblock \showarticletitle{Distillation-Enhanced Graph Masked Autoencoders for Bundle Recommendation}. In \bibinfo{booktitle}{\emph{{SIGIR}}}. \bibinfo{pages}{1660--1669}.
\newblock


\bibitem[Salganik et~al\mbox{.}(2024)]%
        {LARP}
\bibfield{author}{\bibinfo{person}{Rebecca Salganik}, \bibinfo{person}{Xiaohao Liu}, \bibinfo{person}{Jian Kang}, \bibinfo{person}{Yunshan Ma}, {and} \bibinfo{person}{Tat{-}Seng Chua}.} \bibinfo{year}{2024}\natexlab{}.
\newblock \showarticletitle{LARP: Language Audio Relational Pre-training for Cold-Start Playlist Continuation}. In \bibinfo{booktitle}{\emph{{KDD}}}. \bibinfo{publisher}{{ACM}}.
\newblock


\bibitem[Sar~Shalom et~al\mbox{.}(2016)]%
        {listRec}
\bibfield{author}{\bibinfo{person}{Oren Sar~Shalom}, \bibinfo{person}{Noam Koenigstein}, \bibinfo{person}{Ulrich Paquet}, {and} \bibinfo{person}{Hastagiri~P Vanchinathan}.} \bibinfo{year}{2016}\natexlab{}.
\newblock \showarticletitle{Beyond collaborative filtering: The list recommendation problem}. In \bibinfo{booktitle}{\emph{{WWW}}}. \bibinfo{pages}{63--72}.
\newblock


\bibitem[Sheng et~al\mbox{.}(2024)]%
        {AlphaRec}
\bibfield{author}{\bibinfo{person}{Leheng Sheng}, \bibinfo{person}{An Zhang}, \bibinfo{person}{Yi Zhang}, \bibinfo{person}{Yuxin Chen}, \bibinfo{person}{Xiang Wang}, {and} \bibinfo{person}{Tat-Seng Chua}.} \bibinfo{year}{2024}\natexlab{}.
\newblock \showarticletitle{Language Models Encode Collaborative Signals in Recommendation}.
\newblock \bibinfo{journal}{\emph{arXiv preprint arXiv:2407.05441}} (\bibinfo{year}{2024}).
\newblock


\bibitem[Sun et~al\mbox{.}(2023)]%
        {bundleICL}
\bibfield{author}{\bibinfo{person}{Zhu Sun}, \bibinfo{person}{Kaidong Feng}, \bibinfo{person}{Jie Yang}, \bibinfo{person}{Xinghua Qu}, \bibinfo{person}{Hui Fang}, \bibinfo{person}{Yew-Soon Ong}, {and} \bibinfo{person}{Wenyuan Liu}.} \bibinfo{year}{2023}\natexlab{}.
\newblock \showarticletitle{Dynamic In-Context Learning from Nearest Neighbors for Bundle Generation}.
\newblock \bibinfo{journal}{\emph{arXiv preprint arXiv:2312.16262}} (\bibinfo{year}{2023}).
\newblock


\bibitem[Sun et~al\mbox{.}(2022)]%
        {revisit}
\bibfield{author}{\bibinfo{person}{Zhu Sun}, \bibinfo{person}{Jie Yang}, \bibinfo{person}{Kaidong Feng}, \bibinfo{person}{Hui Fang}, \bibinfo{person}{Xinghua Qu}, {and} \bibinfo{person}{Yew~Soon Ong}.} \bibinfo{year}{2022}\natexlab{}.
\newblock \showarticletitle{Revisiting bundle recommendation: Datasets, tasks, challenges and opportunities for intent-aware product bundling}. In \bibinfo{booktitle}{\emph{{SIGIR}}}. \bibinfo{pages}{2900--2911}.
\newblock


\bibitem[Team(2024)]%
        {llama3.1}
\bibfield{author}{\bibinfo{person}{Llama Team}.} \bibinfo{year}{2024}\natexlab{}.
\newblock \bibinfo{title}{The Llama 3 Herd of Models}.
\newblock
\newblock


\bibitem[Touvron et~al\mbox{.}(2023)]%
        {llama2}
\bibfield{author}{\bibinfo{person}{Hugo Touvron}, \bibinfo{person}{Louis Martin}, \bibinfo{person}{Kevin Stone}, \bibinfo{person}{Peter Albert}, \bibinfo{person}{Amjad Almahairi}, \bibinfo{person}{Yasmine Babaei}, \bibinfo{person}{Nikolay Bashlykov}, \bibinfo{person}{Soumya Batra}, \bibinfo{person}{Prajjwal Bhargava}, \bibinfo{person}{Shruti Bhosale}, {et~al\mbox{.}}} \bibinfo{year}{2023}\natexlab{}.
\newblock \showarticletitle{Llama 2: Open foundation and fine-tuned chat models}.
\newblock \bibinfo{journal}{\emph{arXiv preprint arXiv:2307.09288}} (\bibinfo{year}{2023}).
\newblock


\bibitem[Vaswani et~al\mbox{.}(2017)]%
        {transformer}
\bibfield{author}{\bibinfo{person}{Ashish Vaswani}, \bibinfo{person}{Noam Shazeer}, \bibinfo{person}{Niki Parmar}, \bibinfo{person}{Jakob Uszkoreit}, \bibinfo{person}{Llion Jones}, \bibinfo{person}{Aidan~N Gomez}, \bibinfo{person}{{\L}ukasz Kaiser}, {and} \bibinfo{person}{Illia Polosukhin}.} \bibinfo{year}{2017}\natexlab{}.
\newblock \showarticletitle{Attention is all you need}.
\newblock \bibinfo{journal}{\emph{{NeurIPS}}}  \bibinfo{volume}{30} (\bibinfo{year}{2017}).
\newblock


\bibitem[Wei et~al\mbox{.}(2023)]%
        {BundleGT}
\bibfield{author}{\bibinfo{person}{Yinwei Wei}, \bibinfo{person}{Xiaohao Liu}, \bibinfo{person}{Yunshan Ma}, \bibinfo{person}{Xiang Wang}, \bibinfo{person}{Liqiang Nie}, {and} \bibinfo{person}{Tat-Seng Chua}.} \bibinfo{year}{2023}\natexlab{}.
\newblock \showarticletitle{Strategy-aware bundle recommender system}. In \bibinfo{booktitle}{\emph{{SIGIR}}}. \bibinfo{pages}{1198--1207}.
\newblock


\bibitem[Xie et~al\mbox{.}(2010)]%
        {packageRec}
\bibfield{author}{\bibinfo{person}{Min Xie}, \bibinfo{person}{Laks~VS Lakshmanan}, {and} \bibinfo{person}{Peter~T Wood}.} \bibinfo{year}{2010}\natexlab{}.
\newblock \showarticletitle{Breaking out of the box of recommendations: from items to packages}. In \bibinfo{booktitle}{\emph{{RecSys}}}. \bibinfo{pages}{151--158}.
\newblock


\bibitem[Yin et~al\mbox{.}(2023)]%
        {surveyMLLM}
\bibfield{author}{\bibinfo{person}{Shukang Yin}, \bibinfo{person}{Chaoyou Fu}, \bibinfo{person}{Sirui Zhao}, \bibinfo{person}{Ke Li}, \bibinfo{person}{Xing Sun}, \bibinfo{person}{Tong Xu}, {and} \bibinfo{person}{Enhong Chen}.} \bibinfo{year}{2023}\natexlab{}.
\newblock \showarticletitle{A Survey on Multimodal Large Language Models}.
\newblock \bibinfo{journal}{\emph{arXiv preprint arXiv:2306.13549}} (\bibinfo{year}{2023}).
\newblock


\bibitem[Yu et~al\mbox{.}(2022)]%
        {HyperGraph}
\bibfield{author}{\bibinfo{person}{Zhouxin Yu}, \bibinfo{person}{Jintang Li}, \bibinfo{person}{Liang Chen}, {and} \bibinfo{person}{Zibin Zheng}.} \bibinfo{year}{2022}\natexlab{}.
\newblock \showarticletitle{Unifying multi-associations through hypergraph for bundle recommendation}.
\newblock \bibinfo{journal}{\emph{Knowledge-Based Systems}}  \bibinfo{volume}{255} (\bibinfo{year}{2022}), \bibinfo{pages}{109755}.
\newblock


\bibitem[Zamani et~al\mbox{.}(2019)]%
        {music_continuation}
\bibfield{author}{\bibinfo{person}{Hamed Zamani}, \bibinfo{person}{Markus Schedl}, \bibinfo{person}{Paul Lamere}, {and} \bibinfo{person}{Ching-Wei Chen}.} \bibinfo{year}{2019}\natexlab{}.
\newblock \showarticletitle{An analysis of approaches taken in the acm recsys challenge 2018 for automatic music playlist continuation}.
\newblock \bibinfo{journal}{\emph{{TIST}}} \bibinfo{volume}{10}, \bibinfo{number}{5} (\bibinfo{year}{2019}), \bibinfo{pages}{1--21}.
\newblock


\bibitem[Zhang et~al\mbox{.}(2024b)]%
        {speechGPT}
\bibfield{author}{\bibinfo{person}{Dong Zhang}, \bibinfo{person}{Xin Zhang}, \bibinfo{person}{Jun Zhan}, \bibinfo{person}{Shimin Li}, \bibinfo{person}{Yaqian Zhou}, {and} \bibinfo{person}{Xipeng Qiu}.} \bibinfo{year}{2024}\natexlab{b}.
\newblock \showarticletitle{SpeechGPT-Gen: Scaling Chain-of-Information Speech Generation}.
\newblock \bibinfo{journal}{\emph{CoRR}}  \bibinfo{volume}{abs/2401.13527} (\bibinfo{year}{2024}).
\newblock
\showeprint[arXiv]{2401.13527}


\bibitem[Zhang et~al\mbox{.}(2024a)]%
        {GraphTranslator}
\bibfield{author}{\bibinfo{person}{Mengmei Zhang}, \bibinfo{person}{Mingwei Sun}, \bibinfo{person}{Peng Wang}, \bibinfo{person}{Shen Fan}, \bibinfo{person}{Yanhu Mo}, \bibinfo{person}{Xiaoxiao Xu}, \bibinfo{person}{Hong Liu}, \bibinfo{person}{Cheng Yang}, {and} \bibinfo{person}{Chuan Shi}.} \bibinfo{year}{2024}\natexlab{a}.
\newblock \showarticletitle{GraphTranslator: Aligning Graph Model to Large Language Model for Open-ended Tasks}. In \bibinfo{booktitle}{\emph{{WWW}}}. \bibinfo{pages}{1003–1014}.
\newblock


\bibitem[Zhang and Wang(2023)]%
        {prompt-tuning-newsrec}
\bibfield{author}{\bibinfo{person}{Zizhuo Zhang} {and} \bibinfo{person}{Bang Wang}.} \bibinfo{year}{2023}\natexlab{}.
\newblock \showarticletitle{Prompt learning for news recommendation}. In \bibinfo{booktitle}{\emph{{SIGIR}}}. \bibinfo{pages}{227--237}.
\newblock


\bibitem[Zhu et~al\mbox{.}(2023a)]%
        {zhu2023chatgpt}
\bibfield{author}{\bibinfo{person}{Deyao Zhu}, \bibinfo{person}{Jun Chen}, \bibinfo{person}{Kilichbek Haydarov}, \bibinfo{person}{Xiaoqian Shen}, \bibinfo{person}{Wenxuan Zhang}, {and} \bibinfo{person}{Mohamed Elhoseiny}.} \bibinfo{year}{2023}\natexlab{a}.
\newblock \showarticletitle{ChatGPT Asks, BLIP-2 Answers: Automatic Questioning Towards Enriched Visual Descriptions}.
\newblock \bibinfo{journal}{\emph{arXiv preprint arXiv:2303.06594}} (\bibinfo{year}{2023}).
\newblock


\bibitem[Zhu et~al\mbox{.}(2023b)]%
        {minigpt4}
\bibfield{author}{\bibinfo{person}{Deyao Zhu}, \bibinfo{person}{Jun Chen}, \bibinfo{person}{Xiaoqian Shen}, \bibinfo{person}{Xiang Li}, {and} \bibinfo{person}{Mohamed Elhoseiny}.} \bibinfo{year}{2023}\natexlab{b}.
\newblock \showarticletitle{Minigpt-4: Enhancing vision-language understanding with advanced large language models}.
\newblock \bibinfo{journal}{\emph{arXiv preprint arXiv:2304.10592}} (\bibinfo{year}{2023}).
\newblock


\bibitem[Zou et~al\mbox{.}(2023)]%
        {HIDGN}
\bibfield{author}{\bibinfo{person}{Ding Zou}, \bibinfo{person}{Sen Zhao}, \bibinfo{person}{Wei Wei}, \bibinfo{person}{Xian-ling Mao}, \bibinfo{person}{Ruixuan Li}, \bibinfo{person}{Dangyang Chen}, \bibinfo{person}{Rui Fang}, {and} \bibinfo{person}{Yuanyuan Fu}.} \bibinfo{year}{2023}\natexlab{}.
\newblock \showarticletitle{Towards Hierarchical Intent Disentanglement for Bundle Recommendation}.
\newblock \bibinfo{journal}{\emph{{TKDE}}} (\bibinfo{year}{2023}).
\newblock


\end{thebibliography}

\appendix
\begin{figure*}[h]
    \centering
    \includegraphics[width=0.99\linewidth]{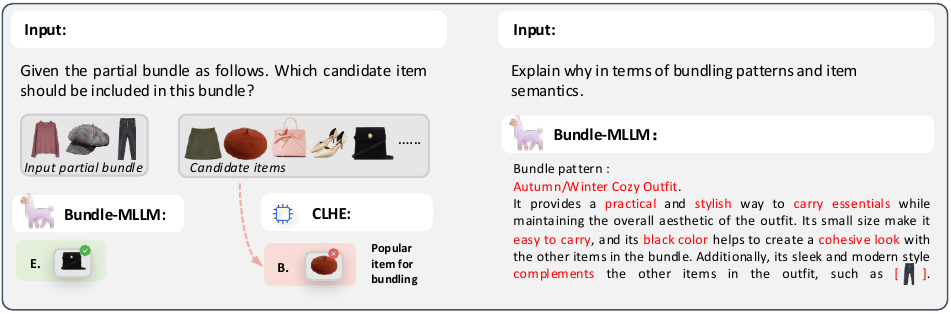}
    \caption{Case studies. Bundle-MLLM recommends the correct candidate item, whereas CLHE tends to recommend popular items for bundling (left). The bundle pattern and item semantic understanding exhibited for product bundling of Bundle-MLLM (right).}
    \label{fig:case}
\end{figure*}

\section{Implementation Details}
\label{appendix:implementation details}
\subsection{Conventional Approaches}
All conventional product bundling methods are initialized using the Xavier method and trained with the Adam optimizer.
We perform a grid search for the learning rate within $\{10^{-2}, 10^{-3}, 10^{-4}\}$ and for L2 regularization within $\{10^{-3}, 10^{-4}, 10^{-5},$ $ 10^{-6}\}$. The embedding size is set to 64.
For adapting conventional approaches for comparison, we shorten the length of candidates as what Bundle-MLLM adopts, and use the computed score for ranking and output the predicted answer. Formally, 
\begin{equation}
    s = [{z_b z_i^\top;i\in \mathcal{I}_c}];\quad \text{answer} = \arg \max s, 
\end{equation}
where $z_b$ and $z_i$ represent the bundle and item representations, respectively, produced by conventional bundling methods. The set of candidate items, $\mathcal{I}_c$, is constructed according to the protocols outlined in Section ~\ref{sec:evaluation_protocols}.

\subsection{LLM-based Approaches}
For all implementations related to LLMs, we use a linear warm-up strategy, where the learning rate starts from 0 and increases to $3\times 10^{-4}$ of the maximum learning rate. 
Each experiment is trained for a maximum of 10 epochs with a batch size of 16. To evaluate the output of the LLM, we valid the answer to compute the ValidRatio metric. Then, all the answer will be transformed to a number to share a unified evaluation with conventional methods. Specifically, we apply $\text{answer} = (t_o - \text{'A'})$, where $t_o$ is the valid output of LLM methods, to obtain the corresponding numerical prediction. 
We do not include any MLLMs as baselines due to their inability to comprehend multiple media content and relational data. 
Nonetheless, the ablation study in Table~\ref{tab:modalities} demonstrates that Bundle-MLLM outperforms \textbf{Text+Media}, an adaptation of MLLMs for product bundling that supports multiple media content as input.

\subsection{Bundle-MLLM}
We select Llama2-7B~\cite{llama2} as the LLM backbone. Other LLMs is also available
More recent released model like Llama3.1~\cite{llama3.1} will be tuned in our future work. 
For a fair comparison, we select Llama2-7B as a baseline and view GPT4+ICL as the strongest LLM-based baseline. 
Notably, our method uses only 1024 samples for training, whereas all conventional baselines are trained with the entire given training dataset.

\section{Case study}
To provide an intuitive exhibition of Bundle-MLLM, we select a specific product bundling question and inquire about the reasoning, as shown in Figure~\ref{fig:case}. 
In this scenario, Bundle-MLLM makes the correct prediction given the input partial bundle (\ie pullover sweater, painter hat, and pencil pants). 
In contrast, CLHE incorrectly chooses a popular fashion item (\ie pumpkin beret), disregarding the complementary requirements. 
To showcase the multimodal semantic understanding, we adapt Bundle-MLLM for chatting, asking for the reasoning in terms of bundling patterns and item semantics. 
Specifically, we use the trained adapters for multimodal feature transformation and instructed version of Llama2 for chatting. 
Bundle-MLLM predicts the bundle pattern as ``Autumn/Winter Cozy Outfit,'' which aligns with the given items and provides comprehensive reasons. These reasons include practical purposes (\ie carrying essentials), aesthetic appeal (\ie black color and cohesive look), and complementary requirements (\ie complementing the other items).
Overall, Bundle-MLLM achieves a significant breakthrough in understanding multimodal semantics from bundle patterns and enhances product bundling tasks.

\end{document}